\documentclass[10pt,journal,compsoc]{IEEEtran}

\usepackage{pgfplots}
\usepackage{xcolor}
\usepackage{textcomp}
\usepackage{graphicx}
\usepackage{booktabs}
\usepackage[utf8]{inputenc}
\usepackage[T1]{fontenc}
\usepackage{pgf-pie}
\usepackage{tikz}
\usetikzlibrary{patterns, matrix, positioning, shapes, arrows, fit}
\usepackage{amsmath}
\usepackage{caption}
\usepackage{siunitx}
\usepackage{ifthen}
\usepackage{afterpage}
\usepackage{bm}
\usepackage{listings}
\usepackage{makecell}
\usepackage{tabularx}
\usepackage{caption}
\usepackage{subcaption}
\usepackage{pgfplotstable}
\usepackage{mdframed}
\usepackage{algorithm}
\usepackage{algpseudocode}
\usepackage{array}
\usepackage{colortbl}
\usepackage{multirow}
\usepackage{enumitem}
\usepackage{threeparttable}
\usepackage{adjustbox}
\usepackage{orcidlink}
\usepackage{hyperref}
\usepackage{amssymb}
\usepgfplotslibrary{groupplots}
\usepackage[table]{xcolor}
\usepackage{ragged2e}
\usepackage{xltabular}
\pgfplotsset{compat=newest}

\newcolumntype{L}{>{\raggedright\arraybackslash}X}

\newif\ifdraft
\draftfalse

\newcommand{\removefig}[1]{}

\tikzset{
  idlepattern/.style={pattern=horizontal lines, pattern color=black},
  stressedpattern/.style={pattern=crosshatch, pattern color=black},
}

\IEEEtitleabstractindextext{
\begin{abstract}
Linux is increasingly deployed in Low Earth Orbit on off-the-shelf (COTS) systems-on-chip (SoCs) never designed for the space radiation environment, and when ionizing particles strike, single-event functional interrupts (SEFIs) crash the kernel without warning. Previous studies measured only board-level cross-sections; which Linux subsystems break, and how a single upset cascades into a full OS-wide failure, remained unknown across architectures under stress and irradiation conditions. We address this gap. We subject three Linux platforms to proton irradiation in the 20–58 MeV range: a Raspberry Pi Zero 2W (40 nm planar ARM Cortex-A53), an NXP i.MX 8M Plus (14 nm FinFET ARM Cortex-A53), and an OrangeCrab ECP5 FPGA hosting a VexRiscV RV32I soft-core (40 nm). Through kernel log forensics, we trace all 133 observed Linux failures, most never previously reported, to their originating kernel handler. Failure profiles split dramatically between nodes: on the two 40 nm platforms, memory-management and driver handlers account for 67–78\% of events, while on the 14 nm SoC approximately 90\% of failures funnel through a single eMMC storage path (56\% filesystem + 34\% driver), demonstrating SEFI-susceptible peripheral can dramatically dictate system reliability. The 14 nm SoC also shows roughly an order-of-magnitude lower Linux-SEFI cross-section, though irradiation geometry and DRAM-exposure differences preclude isolating the process-scaling contribution. 
Reconstructed propagation chains shows faults cascade up to six kernel subsystems before terminal failure in severe events. Rather than blanket redundancy, we pinpoint the exact kernel subsystem boundaries where radiation faults originate, enabling targeted mitigations that harden COTS Linux for orbit.
\end{abstract}

}

\begin{document}
\bstctlcite{IEEEexample:BSTcontrol}

\title{Where Linux Breaks Under Radiation:\\ A Cross-Architecture Kernel-Level Characterization of Proton-Induced Failures in COTS SoCs}
\author{
Saad~Memon\,\orcidlink{0000-0001-5672-7573},
Rafal~Graczyk\,\orcidlink{0000-0003-4570-3431},
Tomasz~Rajkowski\,\orcidlink{0000-0002-6112-5152},
Jan~Swakon\,\orcidlink{0000-0001-9262-7326},
Damian~Wrobel\,\orcidlink{0000-0002-5358-9491},
Sebastian~Kusyk\,\orcidlink{0000-0001-8915-5313},
Seth~Roffe\,\orcidlink{0009-0002-8777-4705},
and Mike~Papadakis\,\orcidlink{0000-0003-1852-2547},~\IEEEmembership{Member,~IEEE}%
\IEEEcompsocitemizethanks{
\IEEEcompsocthanksitem S. Memon and M. Papadakis are with the Interdisciplinary Centre for Security, Reliability and Trust (SnT), University of Luxembourg. 
E-mail: \{saad.memon, mike.papadakis\}@uni.lu
\IEEEcompsocthanksitem R. Graczyk is an Independent Researcher. 
E-mail: rafal@graczyk.io
\IEEEcompsocthanksitem T. Rajkowski is with the National Centre for Nuclear Research (NCBJ), Poland. 
E-mail: tomasz.rajkowski@ncbj.gov.pl
\IEEEcompsocthanksitem J. Swakon, D. Wrobel, and S. Kusyk are with the Institute of Nuclear Physics, Polish Academy of Sciences (IFJ PAN), PL 31342 Krakow, Poland. 
E-mail: \{jan.swakon, damian.wrobel, sebastian.kusyk\}@ifj.edu.pl
\IEEEcompsocthanksitem S. Roffe performed this work while at NASA Goddard Space Flight Center, 8800 Greenbelt Rd, Greenbelt, MD 20771, USA. He is now with Shift5, Inc., Arlington, VA, USA. E-mail: seth.roffe@shift5.io
\IEEEcompsocthanksitem This project has received funding from the European Union's Horizon Europe Research and Innovation programme under Grant Agreement No. 101057511 (EURO LABS), and from the Fonds National de la Recherche in Luxembourg through grant CS20/IS/14689454 (HERA). Work at NASA Goddard Space Flight Center was supported by the NASA Electronic Parts and Packaging Program (NEPP).
}
}

\maketitle
\IEEEdisplaynontitleabstractindextext

\section{Introduction}
\label{sec:introduction}

\begin{figure}[!t]
  \centering
  \includegraphics[width=\columnwidth]{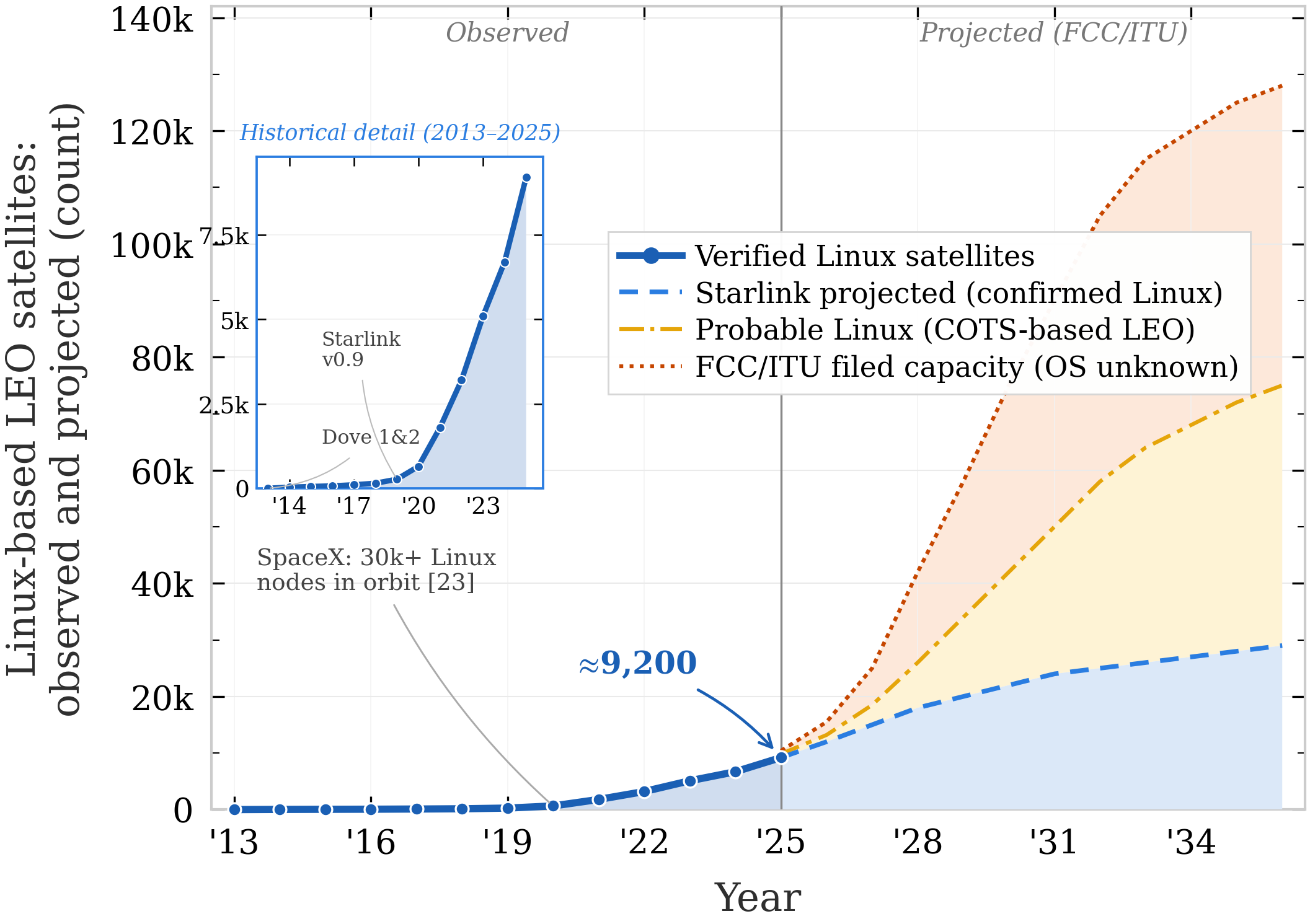}
  \captionsetup{justification=centering}
  \caption{Active Linux-based satellites in LEO: observed counts and projections from FCC and ITU filings.}
  \label{fig:linuxtrend}
\end{figure}

Low-Earth orbit (LEO) avionics are increasingly shifting from radiation-hardened processors paired with real-time operating systems to commercial off-the-shelf (COTS) systems-on-chip (SoCs) running Linux~\cite{nasa_smallsat_avionics_2021,leppinen2017current,miller2023space}, driven by cost, throughput, and time-to-orbit.
Soft real-time Linux has flown on NASA and U.S. Air Force payloads for over a decade~\cite{madden2019challenges,nalepka2001real}; 
SpaceX is reported to operate tens of thousands of Linux-class COTS nodes across Starlink~\cite{Boyle2020Starlink};
OneWeb and Project Kuiper have filed for thousands more by 2030~\cite{fcc_oneweb_2022,fcc_kuiper_2020};
and NASA's Ingenuity Helicopter, a Linux/Snapdragon~801 platform, completed 72 flights against a 5-flight design life on Mars~\cite{balaram2018mars,aagren2025flight}.
Fig.~\ref{fig:linuxtrend} projects this trajectory toward tens of thousands of Linux nodes in orbit within the decade, assuming current trends.

However, COTS SoCs are not designed to tolerate radiation induced failures. 
For instance, a single-event upset (SEU) in any operating register, whether holding computational data or control-path bits, can escalate into a single-event functional interrupt (SEFI)~\cite{1350778,baumann2005radiation,1438282}.
On a Linux platform, the same charge deposit has different consequences depending on \emph{which} bit it flips, a user-space buffer flip is often 
recoverable while a flip in a page table or interrupt vector crashes the running Linux instance outright, making kernel-level vulnerability a first-order reliability concern even at LEO flux.

A SEFI, as defined by JEDEC JESD89A, is a detectable device-level malfunction, lock-up, or reset that recovers without a power cycle.
However, the standard is hardware-scoped, limiting the characterization of software-level manifestations, such as Linux failures that freeze the kernel or corrupt the filesystem state, which require a reboot even when the device remains largely functional.
We therefore use \emph{Linux-SEFI}: a radiation-induced corruption of kernel-mediated state (caused by an SEU or SEFI) after which the running Linux instance cannot recover through its normal reboot path and requires external intervention such as a power cycle.

Linux-SEFI is an increasingly consequential failure mode for space-borne computing.
For instance, mega-constellations re-use identical SoCs and kernel images across thousands of nodes~\cite{rawlins2022death}, so radiation-induced soft errors scale linearly with constellation size~\cite{pfandzelter2023edge}, and any platform-deterministic Linux-SEFI signature manifests as correlated rather than independent outages.

Symmetric multiprocessing (SMP) Linux makes per-board containment harder still: shared caches, interconnects, and memory controllers carry SEFI effects beyond the affected core~\cite{beck2023evaluation,esquer2024single}, as Corley et al.~\cite{corley2022accelerated} demonstrated on a quad-core Raspberry Pi 3B+ where failures appeared on cores not running the workload. 
Software-enforced isolation such as containers and namespaces cannot contain SEFI because every container ultimately executes on the same kernel and physical memory that the SEFI corrupts.

Containment within a system is one challenge; predicting behavior across systems is another. We focus on three hardware axes that shape the Linux-SEFI profile in non-obvious ways.
\emph{(i) Fabrication node} cuts both ways: a smaller node (e.g., 14\,nm FinFET 
vs.\ 40\,nm planar CMOS) shrinks the sensitive volume per bit but packs roughly $4\times$ more SRAM cells into the same area, partially offsetting the per-bit gain at the chip level~\cite{riera2016detailed,Baumann2019Radiation}.
\emph{(ii) Implementation style} changes what an upset becomes: in SRAM-based FPGAs, a single particle can rewrite routing or soft-core logic itself, and the corruption persists until the bitstream is scrubbed, a failure mode absent in fixed-function ASICs~\cite{wilson2023post,quinn2008consequences}. \emph{(iii) ISA choice} (ARM vs.\ RISC-V) routes the same physical upset through different kernel paths, via different trap handlers, interrupt controllers, and firmware mediation, producing different OS-level outcomes from identical faults~\cite{marques2021lock}. Together, these axes determine whether a SEFI manifests as a transient glitch, a persistent fabric corruption, or a correlated fleet-wide outage.

Existing mitigations leave this regime unaddressed.
\emph{(i) TMR} (e.g., Falcon~9 flight COTS computers~\cite{spacex_linux_x86_falcon9}) assumes uncorrelated replica faults.  
Solar particle events (SPEs) break this assumption: proton flux can rise by orders of magnitude during unpredictable bursts~\cite{ecoffet2013overview}, making near-simultaneous Linux-SEFIs across replicas likely, and because every software replica runs on the same kernel, a single SEFI propagates identically to all of them.
TMR is also prohibitive in mass and power for CubeSats. 
\emph{(ii) ECC} leaves pipeline registers, control logic, interconnects, and FPGA configuration unprotected; even strong schemes such as single-error correction (SEC) or double-error correction (DEC) can be overwhelmed by dense multi-bit upsets in a single cycle~\cite{quinn2015software}.
\emph{(iii) Excess COTS compute}, trending an order of magnitude above radiation-hardened parts~\cite{wang2023mars,quinn2015software}, supports redundancy and checkpointing at a low marginal cost~\cite{4522564}, but a SEFI in the shared kernel state propagates identically to every replica, defeating the reliability by the redundancy assumption.

The common point of failure is the kernel itself. 
Hardening kernel state requires mapping which subsystems dominate the fault surface and how upsets cascade between them.
Despite thousands of Linux nodes in orbit, this map does not exist~\cite{4775906}. 
Prior campaigns~\cite{roffe2023single,wilson2023post,esquer2024single,10878262,mojica2018radiation,santini2015reliability,wyrwas2017proton,wyrwas2019proton,jaksch2022debugging} 
report device-level cross-sections on single boards, without tracing Linux 
failures to their originating kernel subsystems~\cite{riera2016detailed,budroweit2021risk} 
or comparing behavior across heterogeneous architectures under matched conditions.

Software fault injection cannot substitute for irradiation: the choice of what to inject is itself a model of the physics being tested, and no campaign can fully reproduce the correlated multi-bit upsets, charge sharing, and secondary nuclear products that a single particle deposits, especially in proprietary microarchitectural states like ARM ISAs.

In this study, we address both gaps: kernel attribution and cross-architecture comparison, 
with proton irradiation (20--58\,MeV) on three Linux-capable devices under test (DUTs) chosen to vary one axis at a time: 
\emph{(i)} Raspberry Pi Zero 2W (quad Cortex-A53, 40 nm planar bulk CMOS, LPDDR2); 
\emph{(ii)} NXP i.MX 8M Plus (quad Cortex-A53, 14 nm FinFET, LPDDR4); 
and \emph{(iii)} OrangeCrab ECP5 SRAM FPGA hosting a VexRiscV RV32I soft-core (DDR3L); 
to introduce a different fabrication technology, implementation style, and ISA.

All three DUTs run without ECC or shielding under matched proton conditions, exposing kernel-level fault response as the sole remaining software-visible failure path.
Because the platforms differ on multiple hardware axes simultaneously (process node, packaging, DRAM, ISA, implementation style), we use them to identify \emph{which kernel paths fail and how failures propagate} across heterogeneous architectures, not to rank platforms or report absolute device-level SEFI cross-sections.

This study addresses two central research questions (RQs):

\textbf{RQ1:} How do Linux-SEFI cross-sections compare across COTS platforms spanning different fabrication nodes, implementation styles, and ISAs under matched proton exposure, and which platform-level factors are consistent with the observed differences?

\textbf{RQ2:} How do proton-induced faults propagate through Linux kernel subsystems into OS-level failures, and what targeted software mitigations do the observed propagation paths motivate?

This work makes the following contributions:
\begin{enumerate}

\item \textbf{First cross-architecture Linux-SEFI measurement under matched proton conditions.} We measure OS-level SEFI cross-sections on three COTS platforms spanning 40\,nm planar CMOS and 14\,nm FinFET, ARMv8-A and RV32I, and hard-core and FPGA soft-core implementations, all irradiated at 20--58\,MeV under matched beam energies, flux, and workload conditions. 
The 14\,nm FinFET platform exhibits approximately one order of magnitude lower Linux-SEFI cross-section and 5--8$\times$ longer mean Linux uptime than the 40\,nm platforms; we attribute this gap to compound platform-level factors (DRAM exposure geometry, packaging, SoC architecture) rather than process scaling alone (Sec.~\ref{sec:evaluation_i},~\ref{sec:dut_imx8mp}; Tables~\ref{tab:proton_test_summary},~\ref{tab:merged_sefi}; Fig.~\ref{fig:sefi_cross_section}).

\item \textbf{Kernel-handler attribution for 133 Linux failures.} We trace every observed failure to its earliest identifiable kernel handler through log forensics. Memory management and device drivers dominate on the 40\,nm platforms (67--78\%), whereas the eMMC storage path dominates on the 14\,nm platform (${\sim}90\%$), evidence that a single peripheral can govern Linux-level reliability more strongly than the primary SoC. We report this as platform-level evidence; SoC architecture, memory technology, and integration also differ across DUTs (Sec.~\ref{sec:evaluation_ii}; Fig.~\ref{fig:linux_sefi_all_dut}; Tables~\ref{tab:sefi_rpi_analysis}--\ref{tab:sefi_ocv2}).

\item \textbf{Cross-subsystem fault propagation and reproducible recovery deadlock.}
Using kernel-trace forensics (timestamps, call traces, error codes), we quantify OS-level failure sequences via amplification $\alpha$ (distinct kernel subsystems reached) and maximum fan-out $\phi_{\max}$ (concurrent causally linked manifestations per node). 
We observed single faults traverse up to six kernel subsystems (memory management, drivers, filesystem, etc) before terminal failure. 
Coincident independent upsets within a run cannot be excluded, but the measured trust-boundary crossings hold under either interpretation. 
In 3 of 133 events, restoring PID\,1 required the same corrupted eMMC subsystem that the fault had disabled, a circular dependency that blocks autonomous recovery (Sec.~\ref{sec:dut2-failure-chains}; Tables~\ref{tab:propagation},~\ref{tab:sefi_imx8mp}).

\end{enumerate}

\section{Testing Methodology}
\label{sec:methodology}

\subsection{Scope of Proton Irradiation Experiment}
\begin{figure*}[htbp]
\centering
\begin{subfigure}[t]{\textwidth}
\centering
\includegraphics[width=\textwidth]{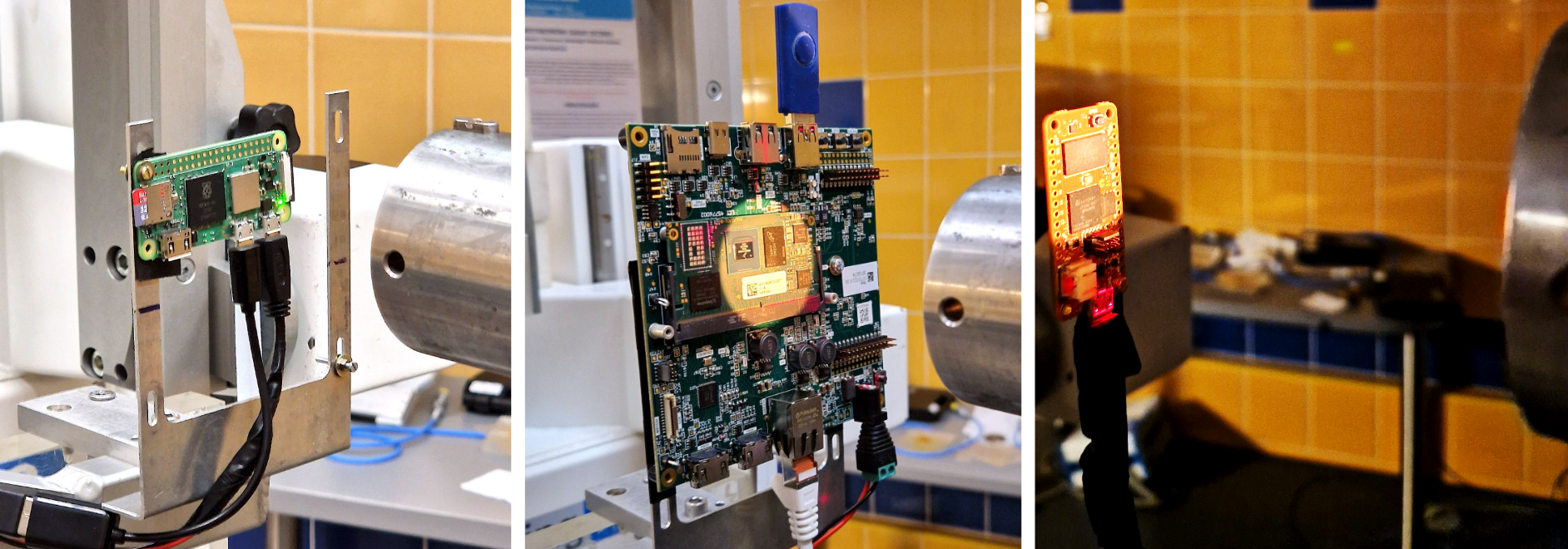}
\end{subfigure}
\captionsetup{justification=centering}
\caption{SoC/software configuration for DUT~1 and DUT~2, along with the monitoring arrangement used in the irradiation chamber.}
\label{fig:dut_setup}
\vspace{2mm}
\end{figure*}

The chosen beam energies (20–58 MeV) employed here evaluate the OS-level Linux-SEFI cross-section. 
Following established COTS soft error screening methodologies~\cite{castillo2011single, y2025towards, pagey2004sefi}, low energy proton irradiation can identify systems with heightened susceptibility to OS level SEUs and SEFIs. 
Fig.~\ref{fig:dut_setup} illustrates the SoC/software configuration for DUT~1 and DUT~2, as well as the monitoring arrangement in our irradiation chamber.

\subsection{Proton Beam Configuration}
\label{sec:proton_rationale}
\begin{figure}[htbp]
\centering
\includegraphics[width=0.45\textwidth]{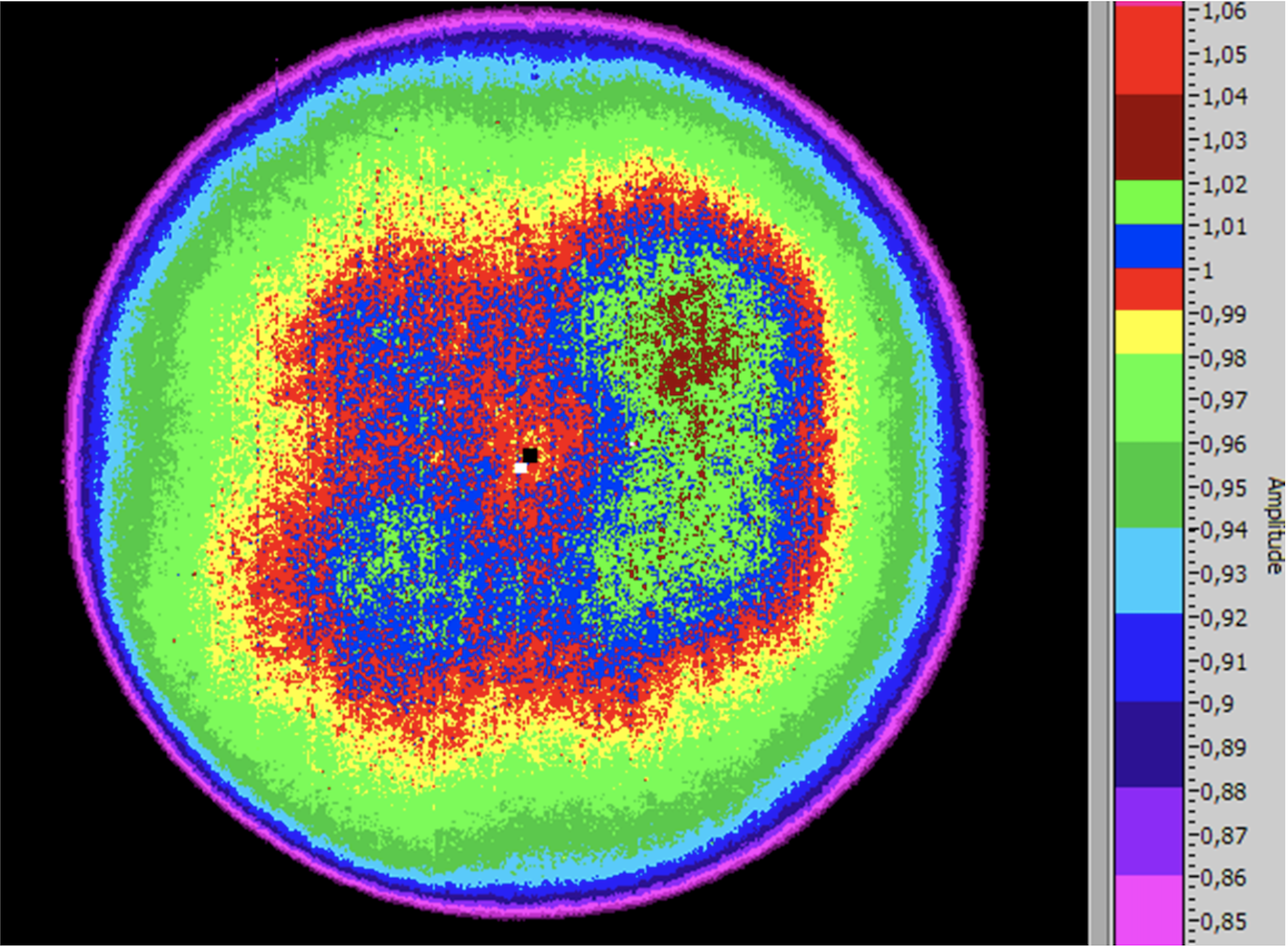}
\captionsetup{justification=centering}
\caption{2D beam profile, normalized intensity.}
\label{fig:beamconf}
\end{figure}

Proton irradiation was performed at the Henryk Niewodniczański
Institute of Nuclear Physics (IFJ PAN) in Kraków, Poland, using the
AIC-144 isochronous cyclotron, which delivers a primary 60~MeV proton
beam. Lower energies in the 20--58~MeV range were obtained with
calibrated degraders, yielding a quasi-monoenergetic spectrum
broadened by energy straggling. Irradiations were carried out in air
at the small-field station, using a 40~mm diameter circular field
with a flat lateral profile. The proton flux ranged from
$3\times10^{6}$ to $1\times10^{8}$~p/cm$^{2}\cdot$s, with an
estimated fluence uncertainty of $\sim$3\%.
Fig.~\ref{fig:beamconf} shows a representative profile acquired
during facility quality assurance.
Fluence and flux were derived from absorbed-dose measurements with a Markus-type parallel-plate ionization chamber in a PMMA
phantom, read out with a PTW UNIDOS reference-class electrometer,
calibrated per the IAEA TRS-398 absorbed-dose
protocol~\cite{musolino2001absorbed}.
Dose-to-fluence conversion used SRIM-2013 stopping powers~\cite{ziegler1985stopping}.
For all DUTs, the depth to active sensitive volumes (0.9--2.3~mm,
depending on the package geometry) was well within the proton range at
the tested energies, ensuring full coverage in both planar and
system-in-package (SiP) configurations.

\subsection{Proton Energy Selection for SEFI Testing}
This campaign targets OS-level failure-mode identification, not extraction of intrinsic device parameters such as threshold LET or Weibull fits, which require heavy-ion testing~\cite{label2009proton, ESCC25100, petersen2011single}. For sub-100\,nm COTS SoCs, nuclear-reaction secondaries from tens-of-MeV protons deposit sufficient charge in modern sensitive volumes to trigger upsets and SEFIs~\cite{label2009proton, weller2010monte, sierawski2009muon, barak2006simple}. We selected 20--58\,MeV for this campaign, consistent with prior SEFI evaluations~\cite{9207996, badia2021comparison, badia2022reliability}.

\begin{figure}[htbp]
    \centering
    \includegraphics[width=0.5\textwidth]{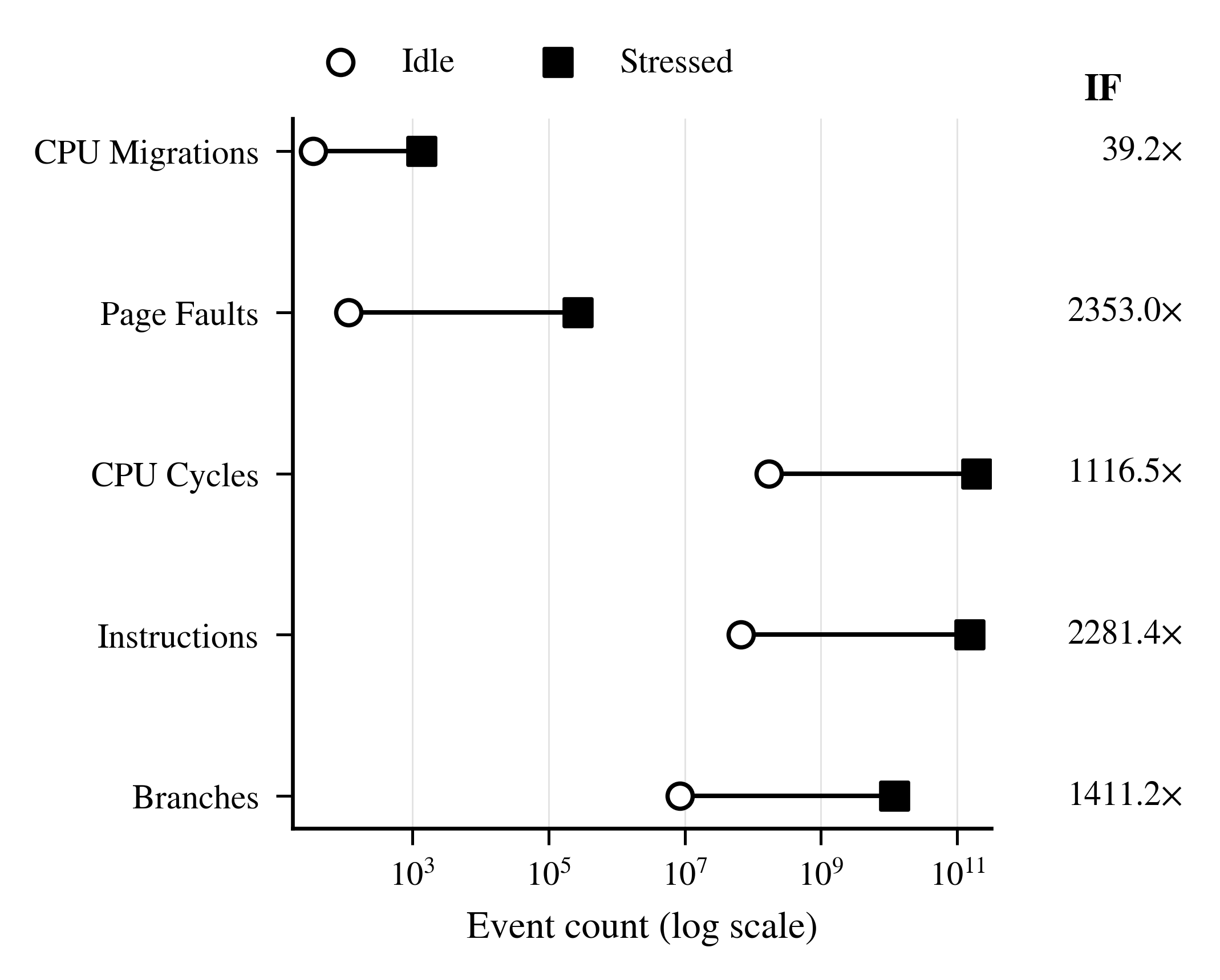}
    \captionsetup{justification=centering}
    \caption{Linux kernel performance metrics: idle vs.\ stressed via \texttt{perf} analysis (IF = Increased Factor).}
    \label{fig:kernel-activity-comparison}
\end{figure}

\subsection{Stress Testing Under Irradiation: Rationale and Workload}
\label{sec:stress_testing}

Accelerated proton testing operates under tight beam-time budgets, requiring a methodology that maximizes failure observability per unit fluence. 
We stress the DUTs to raise the fraction of storage elements holding live, OS-relevant state, thereby increasing the Program Vulnerability Factor (PVF)~\cite{sridharan2009eliminating} at the Linux interface, defined as the conditional probability that an underlying SEU or SEFI propagates into an architecturally visible Linux failure. 
Our reported $\sigma$ values are therefore \emph{workload-conditioned} Linux-SEFI cross-sections, not device-intrinsic SEU cross-sections (Sec.~\ref{sec:evaluation_i}).

We used \texttt{stress-ng}~\cite{king_stressng} to concurrently exercise CPU, memory, and storage subsystems during irradiation using the following configurations:

\begin{center}
\fbox{%
\parbox{0.95\columnwidth}{%
\texttt{stress-ng --cpu 4 --cpu-method fft --vm 1
\hspace*{0em}--vm-bytes 256M --io 1 --hdd 1 --timeout 60s}}}
\end{center}
The FFT workload saturates all cores with working sets exceeding $L_1/L_2$, forcing sustained DRAM traffic; the \texttt{--vm}, \texttt{--io}, and \texttt{--hdd} stressors exercise the page-fault handler, slab allocator, DMA paths, and filesystem code. 
Linux \texttt{perf} measurements confirmed order-of-magnitude increases in page faults, instructions, cycles, context switches, and branch operations versus idle (Fig.~\ref{fig:kernel-activity-comparison}), an activity through which a single SEFI or SEU can propagate across threads via shared kernel state.

\begin{figure}[htbp]
    \centering
    \includegraphics[width=0.5\textwidth]{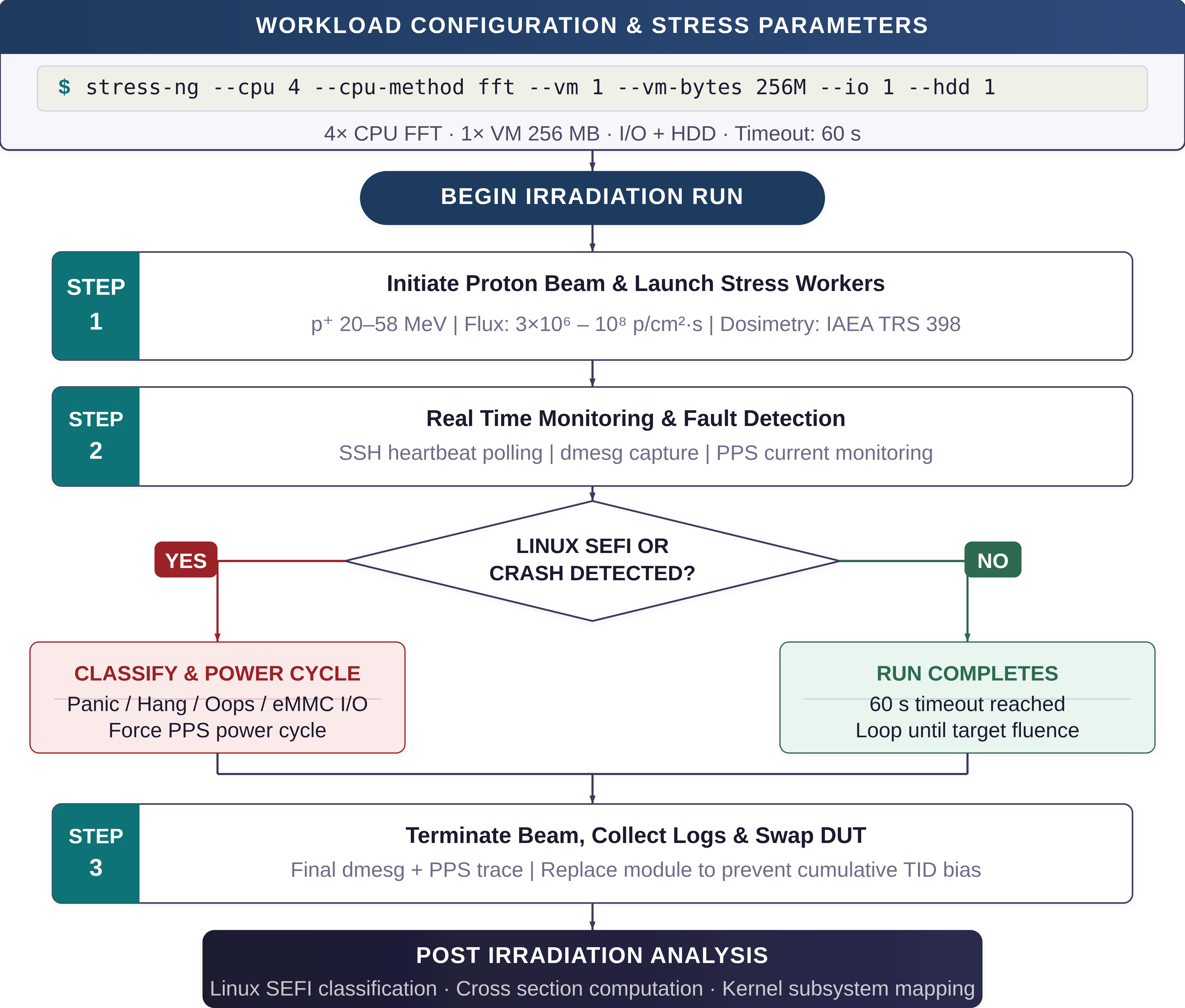}
    \captionsetup{justification=centering}
    \caption{Irradiation protocol employed in the proton beam test campaign.}
    \label{fig:testing_methodology}
\end{figure}

On DUT~1, \texttt{stress-ng} integrity checkers flagged six application-level miscompares (row-hammer, rand-set, gray-code, and $3\times$ Modulo-X; Table~\ref{tab:sefi_rpi_analysis}). These pattern labels are not JEDEC SEFIs, but they corroborate that the stress workload exposes silent corruption under the beam alongside 133 recorded Linux failures, mostly unreported in existing studies.
Fig.~\ref{fig:testing_methodology} shows the irradiation protocol.

While these application-level detections are not SEFIs in the JEDEC sense, their occurrence under the beam corroborates increased fault manifestation under stress and the effectiveness of our testing methodology, alongside 133 recorded Linux failures, mostly unreported in existing studies.
Fig.~\ref{fig:testing_methodology} presents the irradiation
protocol employed throughout the test campaign.

\subsection{Test Infrastructure and Procedures}
\label{sec:test_infrastructure}
Each DUT was powered by an external programmable power supply (PPS) that continuously logged supply current and voltage to a dedicated monitoring computer outside the irradiation chamber.
All DUTs were connected through a wired network to enable SSH-based remote control and log capture, ensuring data was preserved even when a fatal crash prevented the DUT from storing logs locally.
After each targeted fluence, each DUT was replaced with an identical, previously unexposed copy of its compute module. 
This replacement protocol prevents two cumulative degradation mechanisms from biasing the SEFI analysis: total ionizing dose (TID) buildup and displacement damage (DD), in which lattice dislocations from nuclear interactions alter SEE sensitivity~\cite{petersen2011single}.
Analysis of single-event latch-up (SEL), including micro-latch-up, was outside the scope of this study.
Each DUT was booted prior to irradiation, and only Linux-running systems were exposed to avoid boot failures that could bias the Linux-SEFI cross-section.

\subsection{DUT Selection}
\label{sec:dut_selection}

DUT selection targeted hardware suitable for non-critical deployment on small satellites (e.g. CubeSats).
Within that constraint, platforms were ranked by SWaP (Size, Weight, and Power) and per-unit cost as primary criteria, with long-term vendor support ($>$five-year product lifecycle) and an active open-source community as secondary factors.

Three architecturally diverse platforms were selected, Raspberry Pi Zero~2~W, NXP i.MX~8M Plus, and OrangeCrab~v2, spanning two process nodes, two ISAs (ARMv8-A, RV32I), and three packaging types (SiP, FC-BGA, csfBGA). 
This enables a \emph{comparative} (rather than factor-isolated) Linux-SEFI study under matched beam energies and comparable flux. Because the kernel couples all subsystems through shared state, the reported $\sigma$ values characterize OS-level failure under defined beam geometry and platform configuration, not intrinsic per-component susceptibility (cf.\ \S\ref{sec:stress_testing}).

Table~\ref{tab:dutcomparison} summarizes hardware and software specifications. The two ARM DUTs run near-identical 5.15 kernels (5.15.84 and 5.15.77) sharing \texttt{mm}, \texttt{fs}, and driver implementations; the RISC-V DUT runs 5.0.13 on a minimal Buildroot userspace. The 5.0 and 5.15 series differ in scheduler and slab internals, so we target cross-architecture failure-\emph{mode} identification rather than absolute rate comparison.

\begin{table}[t]
\centering
\scriptsize
\renewcommand{\arraystretch}{1.05}
\setlength{\tabcolsep}{2pt}
\caption{System-level specifications of the three DUT platforms used in proton irradiation testing. ECC was not used on any DUT.}
\label{tab:dutcomparison}
\begin{tabularx}{\columnwidth}{@{} l >{\centering\arraybackslash}X >{\centering\arraybackslash}X >{\centering\arraybackslash}X @{}}
\toprule
\textbf{Parameter} & \textbf{RPi Zero 2\,W} & \textbf{NXP i.MX 8M+} & \textbf{OrangeCrab v2} \\
\midrule
Board Type      & SBC (SoC)          & SoM (SoC)          & FPGA Dev Board \\
\midrule
Processor       & BCM2710A1    & i.MX 8M Plus  & VexRiscV (Soft) \\
\midrule
Tech.\ Node     & 40 nm Planar & 14\,nm FinFET & 40\,nm Planar \\
\midrule
CPU             & Cortex-A53   & Cortex-A53    & VexRiscV (synth.) \\
\midrule
ISA             & ARMv8-A (64b) & ARMv8-A (64b) & RV32I (32b) \\
\midrule
Cores           & 4 (quad)     & 4 (quad)      & 1; 24K LUTs \\
\midrule
Frequency       & 1.0\,GHz    & 1.8\,GHz      & 64\,MHz \\
\midrule
RAM             & 512\,MB LPDDR2 & 4\,GB LPDDR4 & 128\,MB DDR3L \\
\midrule
OS              & \makecell[c]{RPi OS Lite\\5.15.84, aarch64} & \makecell[c]{Yocto Linux\\5.15.77, aarch64} & \makecell[c]{Linux 5.0.13\\Buildroot, rv32} \\
\midrule
Power (Peak)    & 0.4--3\,W    & 3--5\,W       & 1.5--6.3\,W \\
\midrule
Package         & \makecell[c]{65$\times$30\,mm\\(SiP on board)} & \makecell[c]{15$\times$15\,mm\\0.5\,mm FC-BGA} & \makecell[c]{10$\times$10\,mm\\0.5\,mm csfBGA} \\
\midrule
\makecell[l]{Irradiated\\Component} & \makecell[c]{SiP: RP3A0-AU\\(BCM2710A1 +\\512\,MB LPDDR2)} & \makecell[c]{i.MX 8M+ SoC\\+ Micron 4\,GB\\LPDDR4 + eMMC} & \makecell[c]{FPGA Fabric\\LFE5U-25F\\24K LUTs} \\
\bottomrule
\end{tabularx}
\end{table}

\subsubsection{DUT~1: Raspberry Pi Zero~2~W}

\begin{figure}[htbp]
\centering
\includegraphics[width=0.5\textwidth]{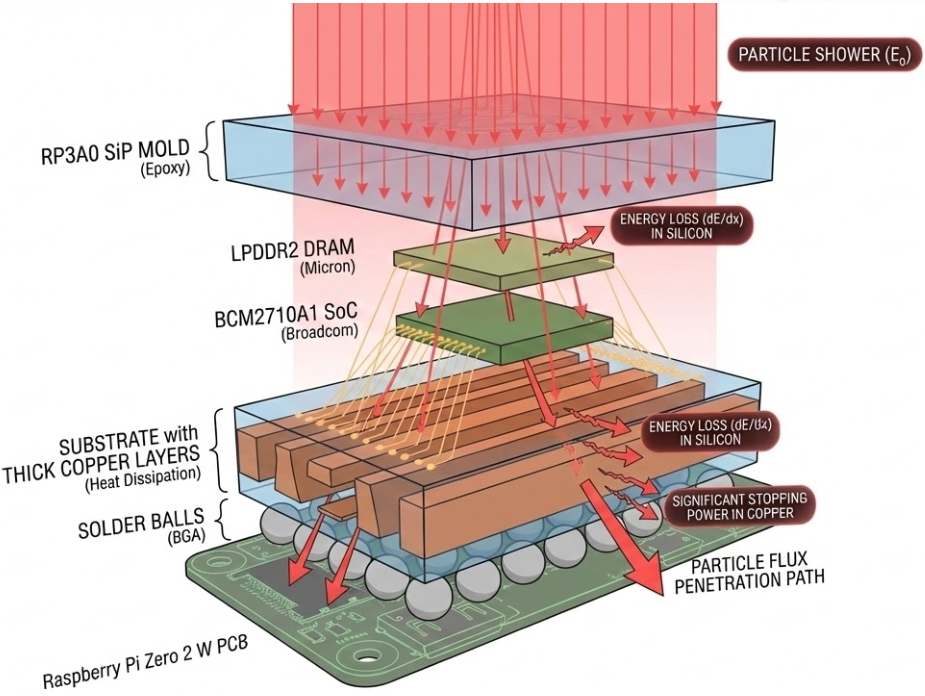}
\captionsetup{justification=centering}
\caption{RP3A0 SiP cross-section: stacked LPDDR2 SDRAM and BCM2710A1 SoC share the proton beam path. At the tested 20--58\,MeV, residual proton energy after the mold and DRAM die is sufficient to deposit charge in the SoC active layer; substrate, BGA, and PCB lie outside the sensitive volume.}
\label{fig:rpi_sip}
\end{figure}

\noindent\textit{Hardware:} The Raspberry Pi Zero~2~W uses an RP3A0-AU SiP integrating a BCM2710A1 quad-core Cortex-A53 SoC with 512~MB LPDDR2 on a single substrate. Its single-core predecessor served as the primary flight computer on NASA's GASPACS 1U CubeSat in LEO~\cite{mojica2018radiation, BuyaRasp86:online}. The SiP's stacked geometry requires incident protons to traverse the full package stack to deposit charge in both die (Fig.~\ref{fig:rpi_sip}); the 20--58\,MeV range exceeds this thickness per \S4.2. Only the SiP was placed in the beam aperture; the microSD root filesystem was kept outside the beam path

\noindent\textit{Software:} Raspbian Linux with automatic root login, granting \texttt{stress-ng} full privileges during irradiation.

\subsubsection{DUT~2: NXP i.MX~8M Plus}

\noindent\textit{Hardware:} This DUT uses a 14,nm FinFET Low Power Compact (LPC) process, which generally reduces charge-collection volume relative to the 40,nm planar CMOS used in the other DUTs. Narasimham et~al.~\cite{narasimham2021scaling} reported approximately one order of magnitude lower per-bit SRAM SER at 16,nm FinFET than at 40,nm planar. A corresponding reduction in system-level SEFI is evaluated in \S\ref{subsec:cross_section_results}. The SoC's inline L2 ECC (768~KB) was disabled to characterize the unprotected baseline.

\noindent\textit{Software:} \texttt{stress-ng} ran in a non-privileged Docker container, which shares the host kernel's memory-mapping paths and is therefore equally exposed to SEUs and SEFIs.

\subsubsection{DUT~3: OrangeCrab v2} \label{sec:dut3_ocv2} \noindent\textit{Hardware:} The OrangeCrab~v2 hosts a Lattice ECP5 FPGA (24K LUT, 40,nm CMOS, csfBGA) configured with a VexRiscV RV32I soft-core at 64,MHz, generated via the LiteX SoC framework and Migen HDL~\cite{kermarrec2020litex}. In contrast to the proprietary ARM Cortex-A53 hard macros of DUT~1 and DUT~2, the entire system, CPU pipeline, Wishbone interconnect, LiteDRAM controller, and peripheral IP, is available as open-source Register Transfer Level (RTL). As an SRAM-based FPGA, the ECP5's configuration memory is itself upset-susceptible; absent scrubbing, configuration SEUs manifest as persistent functional corruption, distinct from the transient logic upsets observed in hard-macro silicon.

\begin{figure}[htbp]
\centering
\includegraphics[width=0.5\textwidth]{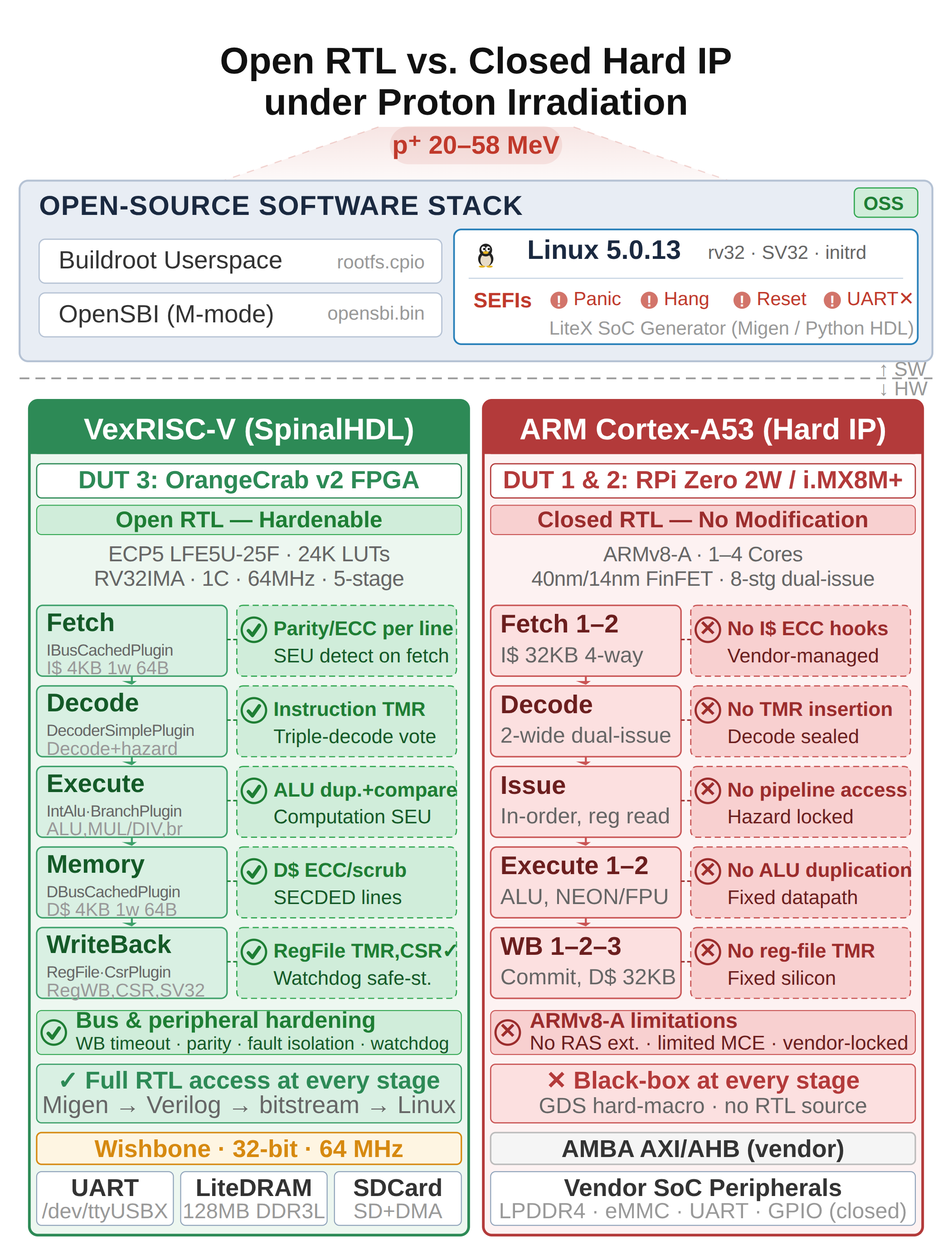}
\vspace{-6pt}
\captionsetup{justification=centering}
\caption{Conceptual comparison of open RTL soft-core (VexRISC-V) and closed hard IP CPU (ARM Cortex-A53) architectures under proton irradiation, highlighting SEFI observability and selective hardening of Linux-SEFI-prone logic.}
\vspace{-6pt}
\label{fig:rtl_hardip}
\end{figure}

This DUT contributes in three ways: (i) Linux-SEFI results for a RISC-V soft-core, complementing the ARMv8-A hard macros; (ii) RTL visibility enabling a test-then-harden workflow via TMR, LiteDRAM ECC, or configuration scrubbing~\cite{6990585} (Fig.~\ref{fig:rtl_hardip}); and (iii) a Linux-capable 64\,MHz core suited to 1U--6U CubeSat supervisory roles, where bitstream reload enables recovery unavailable in fixed silicon~\cite{di2019leveraging}. We test the unmitigated configuration with no TMR and no scrubbing.

Software: A minimal Linux build represents a payload-management role involving sensor I/O, data acquisition, and command handling. Boot artifacts (kernel image, Device Tree Blob (rv32.dtb), OpenSBI firmware, and the compressed root filesystem (rootfs.cpio) on an onboard microSD card remained outside the beam path.

\section{Evaluation I: Linux-SEFI Cross-Section}
\label{sec:evaluation_i}

In this section, we introduce a cross-sectional metric capturing OS-level failure at the Linux interface to address RQ1 (§\ref{sec:introduction}): how SoC architecture, fabrication node, implementation style (ASIC vs.\ FPGA), and ISA shape Linux-level susceptibility under matched irradiation.

\subsection{Defining Linux-SEFI Events}

We define the \emph{Linux-SEFI cross-section} as the workload-conditioned manifestation rate, per unit delivered fluence, of any event that prevents the platform from continuing to execute its Linux workload under matched beam conditions. The metric is platform-level: $\sigma$ integrates every on-die element within the beam aperture, including CPU cores, caches, on-chip SRAM, and, for DUT~3, FPGA configuration memory and programmable logic. The irradiated region is the SoC die (DUT~2 and DUT~3) or SiP (DUT~1, comprising the SoC plus stacked DRAM). Off-die components such as external DRAM on DUT~2/3, PMICs, peripherals, and storage lie outside the beam and contribute only through fault propagation from the irradiated die. Two boundary cases are also classified as Linux-SEFIs: partial failures in which \texttt{stress-ng} crashed while SSH and unrelated processes survived, since the platform no longer executes its workload; and events whose subsystem of origin cannot be uniquely identified from external observation. We make no component-level decomposition; $\sigma$ quantifies OS-visible failure under irradiation of the defined die area, not per-subsystem SEFI cross-section.

\subsection{Linux-SEFI Cross-Section Formulation}
\label{subsec:cross_section_methodology}

The Linux-SEFI cross-section $\sigma$ (cm$^2$) was computed using the standard SEE-test event-per-fluence convention:

\begin{equation}
\sigma = \frac{N_f}{\Phi},
\label{eq:cross_section}
\end{equation}

where $N_f$ is the number of observed Linux-SEFI events and $\Phi$ is the delivered proton fluence (p/cm²), consistent with JEDEC JESD89A and Petersen's SEE-test treatment~\cite{slayman2010jedec,petersen2011single}.

Two independent terms contribute to $\delta\sigma$: counting statistics on $N_f$, and the systematic fluence-delivery uncertainty $\delta\Phi/\Phi = 3\%$. 
For large $N_f$ (typically for a number of errors >50), the Gaussian approximation $\delta N_f \approx 2\sqrt{N_f}$ permits symmetric propagation,
\begin{equation}
\delta\sigma = \sigma \sqrt{\left(\frac{\delta N_f}{N_f}\right)^{\!2} + \left(\frac{\delta\Phi}{\Phi}\right)^{\!2}}.
\label{eq:cross_section_unc}
\end{equation}

This approximation fails at low counts. 
Linux's mean time-to-crash under beam was $\sim$30~s across all three DUTs, so each run terminated on the first SEFI and contributed at most one event; with finite per-point run budgets, several (DUT, energy) cells yielded only 2--5 events. At those counts the Poisson distribution is markedly asymmetric, and symmetric error bars systematically underestimate the upper bound. 
We therefore adopt the exact 95\% Poisson confidence intervals tabulated by Ricker~\cite{RickerWilliam}, with asymmetric bounds $(N_f^-, N_f^+)$ propagated through Eq.~(2) in quadrature with $\delta\Phi/\Phi$ to yield the reported 95\% interval $[\sigma_{\min}, \sigma_{\max}]$ in Table~\ref{tab:proton_test_summary}.

\begin{table}[!htbp]
    \centering
    \caption{Summary of Proton Irradiation Parameters and Linux-SEFI Events Across Three DUTs.}
    \label{tab:proton_test_summary}

    \subfloat[DUT 1: Raspberry Pi Zero 2W (SoC).\label{tab:rpi_pl}]{
    \begin{tabular}{ccccc}
        \toprule
        \textbf{Energy} & \textbf{Runs} & \textbf{Total Linux-} & \textbf{Mean Flux} & \textbf{Total Fluence} \\
        (MeV) & (\textit{n}) & \textbf{SEFI Events} (\textit{f\,/\,n}) & (p/cm$^2$/s) & (p/cm$^2$) \\
        \midrule
        20 & 7  & 7/7   & $3.51 \times 10^{6}$ & $1.08 \times 10^{9}$ \\
        25 & 7  & 7/7   & $3.72 \times 10^{6}$ & $1.22 \times 10^{9}$ \\
        30 & 8  & 8/8   & $3.43 \times 10^{6}$ & $1.15 \times 10^{9}$ \\
        40 & 5  & 5/5   & $3.07 \times 10^{6}$ & $1.06 \times 10^{9}$ \\
        50 & 5  & 5/5   & $2.84 \times 10^{6}$ & $1.05 \times 10^{9}$ \\
        58 & 17 & 17/17 & $4.77 \times 10^{6}$ & $5.98 \times 10^{9}$ \\
        \bottomrule
    \end{tabular}
    }

    \vspace{0.5em}

    \subfloat[DUT 2: NXP i.MX 8M Plus (SoC).\label{tab:imx8mp_pl}]{
    \begin{tabular}{ccccc}
        \toprule
        \textbf{Energy} & \textbf{Runs} & \textbf{Total Linux-} & \textbf{Mean Flux} & \textbf{Total Fluence} \\
        (MeV) & (\textit{n}) & \textbf{SEFI Events} (\textit{f\,/\,n}) & (p/cm$^2$/s) & (p/cm$^2$) \\
        \midrule
        20 & 15 & 15/15 & $1.01 \times 10^{7}$  & $7.59 \times 10^{10}$ \\
        25 & 3  & 3/3   & $1.09 \times 10^{7}$  & $3.57 \times 10^{9}$  \\
        30 & 5  & 4/5   & $1.01 \times 10^{7}$  & $3.55 \times 10^{9}$  \\
        40 & 3  & 2/3   & $1.13 \times 10^{7}$  & $3.51 \times 10^{9}$  \\
        50 & 4  & 3/4   & $1.05 \times 10^{7}$  & $3.51 \times 10^{9}$  \\
        58 & 10 & 5/10  & $1.76 \times 10^{7}$  & $1.61 \times 10^{10}$ \\
        \bottomrule
    \end{tabular}
    }

    \vspace{0.5em}

    \subfloat[DUT 3: OrangeCrab v2 (SoC on FPGA).\label{tab:ocv2_pl}]{
    \begin{tabular}{ccccc}
        \toprule
        \textbf{Energy} & \textbf{Runs} & \textbf{Total Linux-} & \textbf{Mean Flux} & \textbf{Total Fluence} \\
        (MeV) & (\textit{n}) & \textbf{SEFI Events} (\textit{f\,/\,n}) & (p/cm$^2$/s) & (p/cm$^2$) \\
        \midrule
        25 & 5  & 5/5   & $3.21 \times 10^{6}$ & $1.18 \times 10^{9}$  \\
        30 & 9  & 9/9   & $3.12 \times 10^{6}$ & $1.18 \times 10^{9}$  \\
        40 & 7  & 7/7   & $3.22 \times 10^{6}$ & $1.18 \times 10^{9}$  \\
        50 & 4  & 4/4   & $3.39 \times 10^{6}$ & $1.17 \times 10^{9}$  \\
        58 & 32 & 27/32 & $6.86 \times 10^{6}$ & $2.43 \times 10^{10}$ \\
        \bottomrule
    \end{tabular}
    }

\end{table}

\subsection{Measured Linux-SEFI Cross-Sections}
\label{subsec:cross_section_results}

The $\sigma$ vs.\ energy curves (Fig.~\ref{fig:sefi_cross_section}, Table~\ref{tab:merged_sefi}) should not be read as Weibull: under indirect ionization the effective LET is set by the nuclear-reaction secondary spectrum, not the primary proton energy, so the observed energy dependence reflects nuclear-reaction cross-section variation, not a controlled LET sweep~\cite{petersen2011single}.
Weibull extraction requires heavy ions at known LETs. 
The limited platform differentiation at these energies is expected, at low effective LETs, cross-sections cluster near device upset thresholds, with sharper divergence emerging only at saturating heavy-ion LETs.

DUT\,2 falls approximately one order of magnitude below DUT\,1 and DUT\,3 at most energies, while within each platform, the cross-section varies by less than one order of magnitude from 20 to 58\,MeV. 
These cross-section magnitudes are consistent with those reported for comparable COTS processors: Roffe et~al.~\cite{roffe2023pynqz2} measured similar $10^{-8}$\,cm$^{2}$ order cross-sections for 200\,MeV protons on a PYNQ-Z2 (ARM Cortex-A9 with parity ECC), while Cul{\'a}n et~al.~\cite{culan2023polarfire} reported comparable ARM Cortex-A9 results and significantly lower cross-sections for a PolarFire SoC with SECDED ECC under neutron irradiation.
DUT 1 and DUT 3 show their lowest central estimates at 58 MeV, DUT 2 shows a comparably low estimate at 58 MeV but its global minimum at 20 MeV. Overlapping confidence intervals prevent a statistically significant claim of monotonic energy dependence on any platform.
Above ${\sim}20$~MeV, nuclear-reaction secondaries rather than direct ionization by the primary proton are expected to dominate~\cite{petersen2011single}, and the 58~MeV reduction may reflect changes in the secondary spectrum.
The per-DUT analysis below examines what drives the platform separation.

\subsubsection{DUT 1: Raspberry Pi Zero 2 W (40\,nm CMOS)}

DUT\,1 produced the highest cross-sections among the three platforms, ranging from $2.84 \times 10^{-9}$ to $6.95 \times 10^{-9}$\,cm$^{2}$ across 20 to 58\,MeV, with all 49 runs producing Linux-SEFI at all six tested energies.
Central estimates varied by less than a factor of two from 20 to 50\,MeV, indicating weak energy dependence.
At 58\,MeV, the cross-section decreased to $2.84 \times 10^{-9}$\,cm$^{2}$; however, the 95\% Poisson confidence intervals overlap with those at all other energies, making this reduction statistically inconclusive.
The elevated cross-section reflects the aggregate sensitive node count across the BCM2710A1 SoC, LPDDR2, and peripheral subsystems, all fabricated in 40\,nm bulk CMOS.

\subsubsection{DUT 2: NXP i.MX 8M Plus (14\,nm FinFET)}
\label{sec:dut_imx8mp}
DUT\,2 exhibited the lowest cross-sections among all three platforms, ranging from $1.98 \times 10^{-10}$\,cm$^{2}$ at 20\,MeV to $1.13 \times 10^{-9}$\,cm$^{2}$ at 30\,MeV, consistently approximately one order of magnitude below DUT\,1.
The DUT\,2 to DUT\,1 ratio varies with energy: approximately $33\times$ at 20\,MeV ($1.98 \times 10^{-10}$ vs.\ $6.48 \times 10^{-9}$\,cm$^{2}$), narrowing to $6\times$ at 30\,MeV and $9\times$ at 58\,MeV.
This variation reflects DUT\,2's low 20\,MeV central estimate, which carries broad confidence intervals due to the high fluence required to accumulate events.
Uniquely among the three DUTs, not all DUT\,2 runs produced failures: only 5 of 10 at 58\,MeV and 2 of 3 at 40\,MeV yielded events, resulting in broader confidence intervals at those energies.

Two platform-level factors are consistent with the observed gap, although the present experiment cannot separate them:

\textit{(i) Irradiated area and sensitive-volume exposure.} On DUT 1, the irradiated SiP includes both the BCM2710A1 die and the stacked 512 MB LPDDR2 die; on DUT 2, only the i.MX 8M Plus SoC die is within the beam aperture, while the external 4 GB LPDDR4 lies outside it. Because OS-critical kernel state (page tables, slab metadata, process descriptors) is paged from DRAM, on-die DRAM exposure contributes additional sensitive volume that off-die DRAM cannot. We do not compute an area ratio because the per-die active-area dimensions for the BCM2710A1 (DUT 1) are not publicly disclosed; the contribution is therefore attributed qualitatively for Linux-SEFI cross-section.

\textit{(ii) Process technology.}
DUT 2 is fabricated in a 14 nm FinFET process; DUT 1 and DUT 3 use 40 nm bulk planar CMOS.
FinFET devices exhibit reduced charge-collection volumes per transistor compared to CMOS counterparts, which lower the probability of SEU-induced state perturbation~\cite{7398176}. 
However, this per-bit advantage is partially offset by higher bit-cell density at the smaller node, which increases the number of sensitive nodes per unit irradiated area; 
SRAM bit-cell density scaling from 40 nm planar to 14–16 nm FinFET is reported in the range ~4×–6× depending on the cell variant and foundry~\cite{Baumann2019Radiation}.

Scope and limitations:
The measured cross-section reduction reflects the aggregate platform-level behavior. While process technology and irradiation geometry are dominant factors, additional contributions from SoC microarchitecture, DDR subsystem exposure, and power-management circuitry (PMIC) cannot be independently isolated in the present experiment.

\begin{figure}[!t]
    \centering
    \begin{tikzpicture}
        \begin{axis}[
            width=\columnwidth,
            height=0.8\columnwidth,
            xmin=16, xmax=63,
            ymin=3e-11, ymax=2e-7,
            ymode=log,
            xtick={20, 25, 30, 40, 50, 58},
            xticklabels={20, 25, 30, 40, 50, 58},
            minor y tick num=8,
            tick align=inside,
            xtick align=inside,
            ytick align=inside,
            major tick length=3pt,
            minor tick length=1.5pt,
            every tick/.style={black, thin},
            xticklabel style={font=\footnotesize},
            yticklabel style={font=\footnotesize},
            xlabel={Proton Energy (MeV)},
            ylabel={Linux-SEFI Cross-Section, $\sigma$ (cm$^{2}$)},
            xlabel style={font=\small},
            ylabel style={font=\small},
            grid=both,
            major grid style={gray!30, very thin},
            minor grid style={gray!15, ultra thin, dashed},
            legend style={
                at={(0.97,0.96)},
                anchor=north east,
                font=\scriptsize,
                draw=gray!50,
                thin,
                fill=white,
                fill opacity=0.9,
                text opacity=1,
                row sep=1pt,
                inner xsep=4pt,
                inner ysep=2.5pt,
                cells={anchor=west},
                legend columns=1,
            },
            axis lines=box,
            axis line style={thin},
            clip=true,
        ]

        \addplot[
            only marks, mark=*, mark size=2.2pt,
            color=red!80!black, fill=red!80!black,
            error bars/.cd, y dir=both, y explicit,
            error bar style={red!80!black, line width=0.5pt},
            error mark options={rotate=90, mark size=2pt, line width=0.4pt, red!80!black},
        ] table [x=energy, y=sigma, y error minus=errlo, y error plus=errhi] {
            energy  sigma           errlo           errhi
            19.5    6.461141e-09    3.869426e-09    6.831762e-09
            24.5    5.751849e-09    3.444648e-09    6.081784e-09
            29.5    6.979585e-09    4.005880e-09    6.806685e-09
            39.5    4.722327e-09    3.204556e-09    6.328827e-09
            49.5    4.766444e-09    3.234494e-09    6.387952e-09
            57.5    2.841004e-09    1.184343e-09    1.705401e-09
        };
        \addplot[red!80!black, line width=0.5pt, opacity=0.3, no markers, forget plot]
            table [x=energy, y=sigma] {
            energy  sigma
            19.5    6.461141e-09
            24.5    5.751849e-09
            29.5    6.979585e-09
            39.5    4.722327e-09
            49.5    4.766444e-09
            57.5    2.841004e-09
        };
        \addlegendentry{DUT\,1: BCM2710, 40\,nm Bulk CMOS}

        \addplot[
            only marks, mark=square*, mark size=2.2pt,
            color=blue!80!black, fill=blue!80!black,
            error bars/.cd, y dir=both, y explicit,
            error bar style={blue!80!black, line width=0.5pt},
            error mark options={rotate=90, mark size=2pt, line width=0.4pt, blue!80!black},
        ] table [x=energy, y=sigma, y error minus=errlo, y error plus=errhi] {
            energy  sigma           errlo           errhi
            20      1.975842e-10    8.677839e-11    1.291421e-10
            25      8.403361e-10    6.703887e-10    1.624779e-09
            30      1.127714e-09    8.437625e-10    1.748155e-09
            40      5.698006e-10    5.103116e-10    1.481553e-09
            50      8.554320e-10    6.824316e-10    1.653966e-09
            58      3.101352e-10    2.104568e-10    4.156408e-10
        };
        \addplot[blue!80!black, line width=0.5pt, opacity=0.3, no markers, forget plot]
            table [x=energy, y=sigma] {
            energy  sigma
            20      1.975842e-10
            25      8.403361e-10
            30      1.127714e-09
            40      5.698006e-10
            50      8.554320e-10
            58      3.101352e-10
        };
        \addlegendentry{DUT\,2: IMX8MP (14\,nm FinFET)}

        \addplot[
            only marks, mark=diamond*, mark size=2.6pt,
            color=green!60!black, fill=green!60!black,
            error bars/.cd, y dir=both, y explicit,
            error bar style={green!60!black, line width=0.5pt},
            error mark options={rotate=90, mark size=2pt, line width=0.4pt, green!60!black},
        ] table [x=energy, y=sigma, y error minus=errlo, y error plus=errhi] {
            energy  sigma           errlo           errhi
            25.5    4.220478e-09    2.864003e-09    5.656252e-09
            30.5    7.649159e-09    4.241797e-09    6.886054e-09
            40.5    5.923168e-09    3.547247e-09    6.262930e-09
            50.5    3.424658e-09    2.562351e-09    5.308824e-09
            58.5    1.109969e-09    3.774545e-10    5.018856e-10
        };
        \addplot[green!60!black, line width=0.5pt, opacity=0.3, no markers, forget plot]
            table [x=energy, y=sigma] {
            energy  sigma
            25.5    4.220478e-09
            30.5    7.649159e-09
            40.5    5.923168e-09
            50.5    3.424658e-09
            58.5    1.109969e-09
        };
    \addlegendentry{DUT\,3: ECP5, 40\,nm Bulk CMOS}

        \end{axis}
    \end{tikzpicture}
\captionsetup{justification=centering}
\caption{Linux-SEFI cross-section vs.\ proton energy for three COTS platforms (20 to 58\,MeV). Error bars denote 95\% Poisson confidence intervals combined in quadrature with 3\% fluence uncertainty.}
    \label{fig:sefi_cross_section}
\end{figure}

\begin{table}[!htbp]
    \centering
    \caption{Measured Linux-SEFI Cross-Sections with 95\% Poisson Confidence Intervals.}
    \label{tab:merged_sefi}

    \subfloat[DUT 1: Raspberry Pi Zero 2W (SoC).\label{tab:rpi_cc}]{
    \begin{tabular}{cccc}
        \toprule
        \textbf{MeV} & \boldmath$\delta\sigma$ \textbf{(Min) [cm$^2$]} & \textbf{$\bm{\sigma}$ (Observed) [cm$^2$]} & \boldmath$\delta\sigma$ \textbf{(Max) [cm$^2$]} \\
        \midrule
        20 & $2.59 \times 10^{-9}$ & $6.48\times10^{-9}$ & $1.33 \times 10^{-8}$ \\
        25 & $2.30 \times 10^{-9}$ & $5.73\times10^{-9}$ & $1.18 \times 10^{-8}$ \\
        30 & $2.96 \times 10^{-9}$ & $6.95\times10^{-9}$ & $1.38 \times 10^{-8}$ \\
        40 & $1.52 \times 10^{-9}$ & $4.71\times10^{-9}$ & $1.10 \times 10^{-8}$ \\
        50 & $1.53 \times 10^{-9}$ & $4.76\times10^{-9}$ & $1.11 \times 10^{-8}$ \\
        58 & $1.66 \times 10^{-9}$ & $2.84\times10^{-9}$ & $4.54 \times 10^{-9}$ \\
        \bottomrule
    \end{tabular}
    }

    \vspace{0.5em}

    \subfloat[DUT 2: NXP i.MX 8M Plus (SoC).\label{tab:nxp_cc}]{
    \begin{tabular}{cccc}
        \toprule
        \textbf{MeV} & \boldmath$\delta\sigma$ \textbf{(Min) [cm$^2$]} & \textbf{$\bm{\sigma}$ (Observed) [cm$^2$]} & \boldmath$\delta\sigma$ \textbf{(Max) [cm$^2$]} \\
        \midrule
        20 & $1.11 \times 10^{-10}$ & $1.98\times10^{-10}$ & $3.27 \times 10^{-10}$ \\
        25 & $1.70 \times 10^{-10}$ & $8.40\times10^{-10}$ & $2.47 \times 10^{-9}$ \\
        30 & $2.84 \times 10^{-10}$ & $1.13\times10^{-9}$ & $2.88 \times 10^{-9}$ \\
        40 & $5.95 \times 10^{-11}$ & $5.70\times10^{-10}$ & $2.05 \times 10^{-9}$ \\
        50 & $1.73 \times 10^{-10}$ & $8.55\times10^{-10}$ & $2.51 \times 10^{-9}$ \\
        58 & $9.97 \times 10^{-11}$ & $3.10\times10^{-10}$ & $7.26 \times 10^{-10}$ \\
        \bottomrule
    \end{tabular}
    }

    \vspace{0.5em}

    \subfloat[DUT 3: OrangeCrab v2 (SoC on FPGA).\label{tab:ocv2_cc}]{
    \begin{tabular}{cccc}
        \toprule
        \textbf{MeV} & \boldmath$\delta\sigma$ \textbf{(Min) [cm$^2$]} & \textbf{$\bm{\sigma}$ (Observed) [cm$^2$]} & \boldmath$\delta\sigma$ \textbf{(Max) [cm$^2$]} \\
        \midrule
        25 & $1.36 \times 10^{-9}$ & $4.24 \times10^{-9}$ & $9.93 \times 10^{-9}$ \\
        30 & $3.40 \times 10^{-9}$ & $7.63 \times10^{-9}$ & $1.45 \times 10^{-8}$ \\
        40 & $2.38 \times 10^{-9}$ & $5.93 \times10^{-9}$ & $1.22 \times 10^{-8}$ \\
        50 & $8.62 \times 10^{-10}$ & $3.42 \times10^{-9}$ & $8.72 \times 10^{-9}$ \\
        58 & $7.33 \times 10^{-10}$ & $1.11 \times10^{-9}$ & $1.61 \times 10^{-9}$ \\
        \bottomrule
    \end{tabular}
    }

\end{table}

\begin{table*}[!t]
\centering
\caption{Cross-platform DUT comparison of architecture, memory subsystem, and radiation-induced failure characteristics.}
\label{tab:cross-platform}
\renewcommand{\arraystretch}{1.2}
\small
\begin{tabular}{@{}p{2.8cm}p{4.0cm}p{4.0cm}p{4.0cm}@{}}
\toprule
\textbf{Property} & \textbf{DUT1: RPi Zero 2W} & \textbf{DUT2: i.MX 8M Plus} & \textbf{DUT3: OrangeCrab v2} \\
\midrule
ISA / Process & AArch64 / BCM2710 & AArch64 / 14\,nm FinFET & RISC-V / VexRiscV (FPGA) \\
Memory / ECC & LPDDR2 / None & LPDDR4 / None & DDR3L / None \\
Storage & SD card & eMMC (28--40\,nm) & SPI Flash \\
Dominant failure\textsuperscript{a} & DRAM SEU & eMMC SEFI & DDR3L + FPGA fabric SEU \\
Dominant subsystem\textsuperscript{a} & \texttt{mm} (39\%) + \texttt{drivers} (39\%) & \texttt{fs} (56\%) + \texttt{drivers} (34\%) & \texttt{mm} (67\%) \\
Critical mechanism\textsuperscript{b} & VMA pointer corruption $\to$ kernel Oops & eMMC lock $\to$ cascading I/O $\to$ PID\,1 freeze & PTE corruption $\to$ panic \\
\bottomrule
\end{tabular}

\vspace{1mm}

\begin{minipage}{\linewidth}
\raggedright
\footnotesize
\textsuperscript{a}Derived from irradiation test data collected during this campaign.
\textsuperscript{b}Hypothesized fault propagation path inferred from kernel log analysis; definitive confirmation requires per-subsystem fault injection or selective shielding.
\end{minipage}
\end{table*}

\subsubsection{DUT 3: OrangeCrab v2 FPGA (40\,nm CMOS)}
DUT 3, running a VexRiscV soft-core under Linux, exhibited cross-sections from $1.11 \times 10^{-9}$ to $7.63 \times 10^{-9}$ cm², the same order of magnitude as DUT 1. At 30 MeV the DUT 3 central estimate ($7.63 \times 10^{-9}$ cm²) is in fact slightly above the DUT 1 value ($6.95 \times 10^{-9}$ cm²), and the 95\% Poisson intervals overlap; the two values are statistically indistinguishable, not "identical". At 58 MeV, DUT 3 ($1.11 \times 10^{-9}$ cm²) is below DUT 1 ($2.84 \times 10^{-9}$ cm²) with overlapping intervals. Of 57 runs, 52 (91\%) produced Linux-SEFI; at 58 MeV, 27 of 32 runs failed (84\%).

The ECP5 FPGA relies on SRAM-based configuration memory for routing, look-up tables, and the soft-core processor itself; a single upset here creates a persistent fault mechanism distinct from transient upsets in ASIC registers or caches~\cite{morgan2005seu}. 
Despite this vulnerability, DUT\,3's aggregate cross-section remained within the same order of magnitude as DUT\,1. 
However, this similarity does not imply a shared dominant failure mechanism, DUT\,3 may fail primarily from configuration memory upsets, whereas DUT\,1's failures may stem from DRAM or cache faults. 
Decomposing these contributions requires per-subsystem instrumentation beyond the present campaign's scope.

For the cross-DUT comparison required by RQ1, DUT 3 confounds two axes that DUT 1 and DUT 2 do not: implementation style (SRAM-FPGA vs hard macro) and ISA (RV32I vs ARMv8-A), with synthesis, routing, and place-dependent layout variability on top. We accordingly treat the DUT 3 cross-section as an unmitigated soft-core baseline against which future RTL-level hardening~\cite{wilson2021neutron} can be measured, and exclude it from the quantitative DUT 1-vs-DUT 2 process-node comparison.

\section{Evaluation: SEFI-Induced Linux Failures}
\label{sec:evaluation_ii}

\begin{figure*}[!t]
\centering
\begin{minipage}[b]{0.32\textwidth}
    \centering
    \begin{tikzpicture}[baseline=(current bounding box.south)]
        \pie[
            radius=2.0,
            color={blue!70, orange!80, green!70!black, red!70},
            explode={0.1,0,0,0},
            text=pin,
            pin edge={->,>=stealth,shorten <=1pt,shorten >=1pt},
            pin distance=3mm,
            before number=\scriptsize\color{black},
            after number=\%
        ]{
            39/mm, 39/drivers, 16.45/kernel, 5.55/fs
        }
    \end{tikzpicture}
    \vspace{0.5em}\\
    \small (DUT1: rpi02w)
\end{minipage}
\hfill
\begin{minipage}[b]{0.32\textwidth}
    \centering
    \begin{tikzpicture}[baseline=(current bounding box.south)]
        \pie[
            radius=2.0,
            color={blue!70, orange!80, red!70, purple!70},
            explode={0,0,0.1,0},
            text=pin,
            pin edge={->,>=stealth,shorten <=1pt,shorten >=1pt},
            pin distance=3mm,
            before number=\scriptsize\color{black},
            after number=\%
        ]{
            5/mm, 34/drivers, 56/fs, 5/block
        }
    \end{tikzpicture}
    \vspace{0.5em}\\
    \small (DUT2: imx8mp)
\end{minipage}
\hfill
\begin{minipage}[b]{0.32\textwidth}
    \centering
    \begin{tikzpicture}[baseline=(current bounding box.south)]
        \pie[
            radius=2.0,
            color={blue!70, green!70!black, red!70},
            explode={0.1,0,0},
            text=pin,
            pin edge={->,>=stealth,shorten <=1pt,shorten >=1pt},
            pin distance=3mm,
            before number=\scriptsize\color{black},
            after number=\%
        ]{
            67/mm, 22/kernel, 11/fs
        }
    \end{tikzpicture}
    \vspace{0.5em}\\
    \small (DUT3: ocv2)
\end{minipage}

\vspace{1em}

\begin{tikzpicture}
    \node at (0,0) {
        \fbox{%
            \parbox{0.75\textwidth}{%
                \centering
                \small
                \tikz \draw[fill=blue!70] (0,0) rectangle (0.8em,0.8em); \,mm \quad
                \tikz \draw[fill=orange!80] (0,0) rectangle (0.8em,0.8em); \,drivers \quad
                \tikz \draw[fill=green!70!black] (0,0) rectangle (0.8em,0.8em); \,kernel \quad
                \tikz \draw[fill=red!70] (0,0) rectangle (0.8em,0.8em); \,fs \quad
                \tikz \draw[fill=purple!70] (0,0) rectangle (0.8em,0.8em); \,block
            }%
        }
    };
\end{tikzpicture}

\captionsetup{justification=centering}
\caption{Vulnerability shares for Linux kernel components across different DUTs: Memory management (mm) dominates in DUT1 (39\%) and DUT3 (67\%), while filesystems (fs) account for 56\% in DUT2.}
\label{fig:linux_sefi_all_dut}
\end{figure*}
 
Linux does not degrade gracefully under radiation.
In our experiments, localized physical upsets propagated into OS-wide irrecoverable states through implicit trust relationships between kernel layers.
On DUT\,2, a proton-induced eMMC controller lockup traversed six kernel abstraction boundaries before terminating in a circular dependency: every recovery path required the very subsystem that had failed.
This sequence recurred three times with identical symptom signatures, indicating a reproducible architectural failure mode rather than an isolated event.

Cross-platform proton irradiation of three COTS devices spanning two ISAs (AArch64, RISC-V), three memory technologies (LPDDR2, LPDDR4, DDR3L), and three storage architectures (SD card, eMMC, SPI flash), all without ECC, reveals that failure profiles diverge sharply by hardware architecture, yet the underlying vulnerability mechanism is consistent across all tested platforms: Linux's layered trust model assumes lower-layer reliability, and radiation-induced SEFIs violate these assumptions at multiple layers simultaneously.
Our test logs also recorded failures in services that were not exercised during irradiation (HDMI output, Wi-Fi, USB logging). 
Since future crewed missions (e.g., LUCCI~\cite{lucci_mission}) will require peripheral interfaces such as Bluetooth and USB for crew interaction, driver hardening, not merely driver removal, is necessary for operational deployments.
We observe failures through software-visible channels: \texttt{dmesg},
\texttt{journalctl}, userspace signals (\texttt{SIGSEGV}, \texttt{SIGBUS},
\texttt{SIGILL}), driver timeouts (\texttt{-ETIMEDOUT}), I/O errors
(\texttt{-EIO}), watchdog state, and kernel taint flags. Root-cause
attributions combine bit-pattern forensics, temporal correlation, and
architectural consistency; where ambiguity remains, we state it explicitly.

\subsection{Cross-Platform Failure Profiles}
\label{subsec:cross-platform}

Table~\ref{tab:cross-platform} summarizes architectural properties and
radiation-induced failure characteristics across all three DUTs. Because all
devices received identical proton energy and flux, divergent failure profiles
isolate hardware architecture as the primary independent variable.
Fig.~\ref{fig:linux_sefi_all_dut} summarizes the kernel-subsystem vulnerability shares per DUT, motivating the per-platform failure-profile analysis that follows.
The following provides per-DUT analysis. 

\textit{(i) DUT 1 (RPi Zero 2W):} Without ECC, LPDDR2 exposes kernel data structures to direct bit-flip corruption.
Failures are split equally between \texttt{mm} and \texttt{drivers} (39\% each), driven by pointer corruption in VMA structures and DMA buffers. The VideoCore~IV GPU firmware mediates thermal monitoring, DVFS, and display management through a single mailbox interface, making it a hidden single point of failure: one VC4 SEFI simultaneously disabled three ostensibly independent subsystems in our tests. 
To our knowledge, no prior radiation effects study has identified this firmware-mediated single-point-of-failure mode in the context of COTS SoCs.

\textit{(ii) DUT 2 (i.MX 8M Plus):} Although the SoC's \SI{14}{nm} FinFET process reduces per-transistor cross-sections, the eMMC controller dominated system-level failure events, with 56\% of all observed failures manifesting through the filesystem subsystem.
This observation indicates that peripheral controllers, even when fabricated at more radiation-tolerant process nodes, can dominate system-level failure when the kernel depends on them for root filesystem, page cache, swap, and logging services.

\textit{(iii) DUT 3 (OrangeCrab v2):} SRAM-based FPGA fabric introduces a vulnerability class absent from ASIC platforms: proton upsets in configuration memory alter synthesized logic, including the memory controller, interrupt dispatcher, and bus arbiter, producing failure modes with no software-level precursor. The 67\% \texttt{mm} dominance reflects PTE corruption from both DDR3L upsets and FPGA fabric upsets corrupting LiteDRAM controller state. 
On reconfigurable platforms, the boundary between hardware fault and software fault becomes ambiguous, as a configuration memory upset manifests identically to a logic design error from the software's perspective, challenging conventional hardware/software fault isolation assumptions.

These divergent profiles share a common thread: in every case, the kernel's
trust in a lower-layer component, memory, storage, or logic fabric, enabled
a localized upset to propagate unchecked across abstraction boundaries. The
next subsection formalizes these propagation paths.

\subsection{Failure Propagation Chains}
\label{subsec:propagation}

\begin{table}[!b]
\centering
\caption{Failure propagation chain characteristics.}
\label{tab:propagation}
\small
\resizebox{\columnwidth}{!}{%
\begin{tabular}{@{}llccl@{}}
\toprule
DUT  & Chain            & $\boldsymbol{\alpha}$ &
$\boldsymbol{\phi_{\max}}$ & Terminal State \\
\midrule
DUT 1 & VMA corruption   & 4 & 2 & Panic + DRM failure \\
DUT 1 & VC4 SEFI         & 4 & 3 & Thermal/DVFS/HDMI blind \\
DUT 2 & eMMC SEFI        & 6 & 6 & PID\,1 freeze \\
DUT 2 & Docker fault     & 4 & 1 & Service permanently failed \\
DUT 3 & PTE corruption   & 4 & 1 & Use-after-free risk \\
DUT 3 & Config.\ memory  & 3 & 1 & Immediate panic \\
\bottomrule
\end{tabular}%
}
\end{table}

To understand how a single physical upset escalates into a system-wide failure, we model each observed failure sequence as a \emph{hypothesized} propagation chain $\mathcal{C} = (f_0, \ldots, f_n)$, where $f_0$ is the inferred initial upset and each transition $f_i \to f_{i+1}$ crosses a kernel abstraction boundary. Chains are reconstructed from kernel log forensics (timestamps, call traces, error codes), assuming a single root cause where log evidence permits. 
However, temporally coincident failures from independent upsets within the same beam exposure interval cannot be excluded.

Two metrics characterize chain severity: the \emph{amplification factor} $\alpha$, the number of distinct kernel subsystems the chain traverses, and the \emph{fan-out} $\phi_{\max}$, the number of independent failure manifestations triggered at any single node. 
We define chains with $\alpha > 2$ and $\phi > 1$ as \emph{cascading failures}; every critical-severity outcome in our tests resulted from such a cascade under the single-root-cause interpretation. 
Table~\ref{tab:propagation} summarizes the chains observed across all DUTs.

Three patterns emerge across all platforms: (1)~the longest chains ($\alpha \geq 4$) produce exclusively Critical-severity Linux failures; (2)~fan-out $\phi > 1$ generates apparently independent symptoms that are \emph{consistent with} single root causes, complicating post-hoc root-cause attribution in fielded systems; and (3)~inferred propagation depth correlates with the number of trust boundaries crossed, not with physical SEFI severity. These patterns hold whether individual chains are interpreted as single-cause cascades or decomposed into fewer-layer sub-chains with independent co-temporal upsets, because the trust-boundary crossing mechanism is the same in either case.

\subsubsection{DUT 1: Raspberry Pi Zero 2 W Failure Chains}
Chain~A ($\alpha{=}4$, $\phi_{\max}{=}2$): the log 
evidence is consistent with a single-bit DRAM upset that corrupted a VMA \texttt{vm\_next} pointer to non-canonical address
\texttt{0x0001000000000000} (consistent with bit~48 flipped), producing a Level-0 translation
fault during \texttt{exit\_mmap()}, recursive Oops, kernel panic, and
persistent DRM corruption surviving warm reset. 
Chain~B ($\alpha{=}4$, $\phi_{\max}{=}3$): a probable VC4 firmware SEFI simultaneously disabled thermal monitoring, DVFS regulation, and HDMI output, three failures presenting as independent events in \texttt{dmesg} but hypothesized to share a single root cause through the shared mailbox interface.
Table~\ref{tab:sefi_rpi_analysis} details these failures.

\subsubsection{DUT 2: i.MX 8M Plus Failure Chains}
\label{sec:dut2-failure-chains}
Chain ($\alpha{=}6$, $\phi_{\max}{=}6$) progresses from storage SEFI to full-system deadlock. The induced upset locked the eMMC controller in a permanent busy state (L0). The \texttt{mmc} driver detected the hang and executed its full recovery sequence, including cache flush followed by hardware reset via \texttt{RST\_n}, but every attempt returned \texttt{-ETIMEDOUT}; the controller remained unresponsive (L1). EXT4's \texttt{JBD2} journaling thread, finding its commit I/O had failed, aborted the journal and forced a read-only remount of the root filesystem (L2). We hypothesize that the page cache still held executable text segments from incomplete pre-SEFI transfers; the three subsequent \texttt{SIGILL} traps and three \texttt{SIGSEGV} faults (L3) are consistent with demand-paging of corrupted code and data.

Among the victims was PID~1 itself: systemd caught SIGSEGV and attempted a core dump, but with the filesystem read-only the write failed, causing PID~1 to enter pause(), ceasing all process management (L4). Recovery required loading reboot.target from the same unresponsive storage, which returned -EIO (L5). The result was an irresolvable circular dependency: PID~1 needed a functioning filesystem to orchestrate recovery, while filesystem recovery needed a functioning PID~1.

A subsidiary chain ($\alpha{=}4$, $\phi_{\max}{=}1$) targeted the Docker container runtime: the Go runtime received SIGBUS during demand paging of executable text, and the panic handler itself triggered a second SIGBUS when formatting the error message required paging from the failed storage — a double fault eliminating the last opportunity for graceful error reporting. Four subsequent restart attempts failed, exhausting systemd's start-limit threshold and permanently marking the service as failed. Table~\ref{tab:sefi_imx8mp} details these failures.

\subsubsection{DUT 3: OrangeCrab v2 Failure Chains}
Chain~A ($\alpha{=}4$, $\phi_{\max}{=}1$): the bit pattern of the 
corrupted PTE (\texttt{0x10444cd7}) is consistent with a DDR3L upset that corrupted a page table entry, resulting in an invalid value with reserved bits set and a non-existent physical page number, violating RISC-V Sv32 structural invariants. This produced cascading page map errors and a use-after-free risk.
Chain~B ($\alpha{=}3$, $\phi_{\max}{=}1$): an FPGA configuration memory upset corrupted the PLIC interrupt controller, vectoring the CPU to an invalid ISR address. The resulting exception occurred in non-preemptible interrupt context, triggering an immediate kernel panic with no filesystem flush or userspace notification.
Table~\ref{tab:sefi_ocv2} details these failures.

\subsection{Linux Kernel Vulnerability Analysis}
\label{subsec:linux-vuln}
Having demonstrated propagation empirically, we now identify the three specific design assumptions that enabled these cascades:

\emph{(1)~PID 1 non-restartability creates circular recovery dependency:}
The kernel panics if PID~1 terminates, making the init process irreplaceable at runtime. This deadlock occurred three times on DUT~2 with identical signatures, each time transforming the init system from recovery coordinator into recovery blocker. A watchdog mechanism independent of both PID~1 and the storage subsystem, such as a minimal kernel-resident reset path or a hardware watchdog timer, is therefore needed to break this dependency cycle.

\emph{(2)~Memory correctness is assumed between accesses:}
In its default production configuration, the kernel performs no continuous integrity verification on pointer-linked data structures such as \texttt{vm\_area\_struct} lists, nor does the MMU validate page-table entry consistency between translations, both assume hardware returns correct data.
On DUT 1, a single-bit VMA pointer corruption remained latent across multiple list traversals before \texttt{exit\_mmap()} dereferenced the invalid pointer; on DUT 3, corrupted PTEs caused incorrect MMU translations that went undetected until page teardown exposed the inconsistency.
ECC provides incomplete protection against such faults: standard SECDED codes correct only single-bit errors, while MBUs from a single particle strike can exceed correction capacity or, in special cases, alias to valid code words, evading detection entirely~\cite{6217302}.
Each pointer dereference thus assumes memory correctness without verification, making linked kernel data structures a latent vulnerability surface under radiation.
Kernel-level runtime integrity mechanisms independent of ECC, such as software-based pointer tagging or periodic structure validation, remain an open architectural challenge for radiation-exposed Linux systems.

\emph{(3)~Firmware health is neither observable nor verified:}
On DUT 1, the kernel interfaces with the RPi's VC4 firmware through a mailbox channel with no health monitoring, no timeout, and no response validation, treating all firmware responses as authoritative regardless of firmware state.
On DUT 1, a single VC4 SEFI silently corrupted thermal, clock, and power-domain management simultaneously, with nothing at the kernel level to distinguish firmware failure from independent subsystem failures.
Firmware, therefore, operates as an unmonitored single point of failure whose blast radius spans otherwise independent subsystems, a vulnerability that has received limited attention in radiation-effects literature.
Kernel-firmware interfaces with response plausibility checks and hardware-independent watchdog timeouts are needed to break this silent failure mode.

\subsection{Discussion: Mitigation Effectiveness and Limitations}
\label{subsec:mitigation}

\begin{table*}[!t]
\centering
\caption{Mitigation effectiveness against empirically observed failure modes.}
\label{tab:mitigation}
\small
\begin{tabular}{@{}llll@{}}
\toprule
\textbf{Mitigation} & \textbf{Coverage} & \textbf{Overhead} & \textbf{Primary Limitation} \\
\midrule
Hardware ECC (SECDED) & Single-bit DRAM SEUs & ${\sim}12\%$ memory bandwidth & Multi-bit upsets bypass detection \\
\texttt{PANIC\_ON\_OOPS} & Post-Oops corruption propagation & Availability reduction & Does not prevent initial upset \\
Memory scrubbing & Latent DRAM corruption & CPU + memory bandwidth & Vulnerability window $=$ scrub interval \\
dm-verity & Storage data corruption & ${\sim}10\%$ read latency & Requires read-only root filesystem \\
FPGA TMR & Soft-core logic upsets & $3\times$ FPGA resources & Memory subsystem unprotected \\
External watchdog & System hangs / PID\,1 freeze & Board complexity & Cannot prevent data loss before reset \\
Read-only root + tmpfs & Persistent state corruption & Volatile state lost on reboot & Complex state management \\
\bottomrule
\end{tabular}
\end{table*}

In Table~\ref{tab:mitigation}, we evaluate candidate countermeasures against the
observed failure modes. Each mitigation addresses a specific failure class but
introduces its own limitation: hardware ECC (SECDED) corrects single-bit DRAM SEUs but
cannot prevent storage SEFI or multi-bit upsets;
\texttt{PANIC\_ON\_OOPS} halts corruption propagation but trades availability
for safety; dm-verity detects storage corruption but requires read-only root;
TMR guards FPGA logic at $3\times$ resource cost but leaves memory
unprotected; external watchdogs bound hang duration but cannot prevent data
loss before reset.

Defense-in-depth combining ECC, kernel hardening (e.g. page poisoning), filesystem isolation (read-only root, dm-verity), and independent hardware reset is necessary, with the specific combination
determined by the platform's dominant failure archetype
(Table~\ref{tab:cross-platform}). 
The deeper architectural lesson extends beyond any mitigation stack: Linux's trust-based layering assumes reliability is inherited upward from hardware, but radiation violates this assumption at multiple layers simultaneously. Designing radiation-tolerant Linux systems therefore requires not merely adding protections atop the existing architecture, but restructuring trust relationships replacing implicit assumptions with explicit verification at each abstraction boundary.

Since drivers account for 34--39\% of Linux-SEFI events on DUT\,1 and DUT\,2, isolating them from kernel address space is a compelling mitigation largely unexplored in the radiation-effects context. 
Herder et~al.~\cite{herder2009fault} showed driver faults dominate monolithic kernel failures and that running drivers as isolated user-space processes markedly improves reliability.
Chou et al. \cite{10.1145/502034.502042} found that the \texttt{drivers} exhibited three to seven times the error rate of other Linux kernel directories.

Tables~\ref{tab:sefi_rpi_analysis}, \ref{tab:sefi_imx8mp}, and \ref{tab:sefi_ocv2} present the complete per-platform Linux-SEFI failure classification, including fault details traced to kernel source handlers, root-cause hypotheses with cascading impact assessments, and proposed mitigation strategies.
Root-cause attributions are inferred from kernel log forensics; where ambiguity remains, we state it explicitly.
Here, we classify failures by \emph{recovery cost}, i.e., the intervention needed to restore nominal operation, rather than by physical fault type.
\textit{Critical:} irrecoverable without a hardware power cycle (peripheral SEFI, kernel panic in interrupt context).
\textit{High:} persistent degradation requiring a reboot (filesystem journal
abort, read-only remount). 
\textit{Medium-to-Low:} functional impairment with degraded continuation or recoverable warnings without functional loss (e.g., kernel Oops, taint flag accumulation).
This recovery-cost taxonomy structures the cross-platform comparison that follows.

\section{Related Work}
\label{sec:related_work}

Table~\ref{tab:related_work} maps prior Linux-under-radiation studies across four capabilities. Three patterns emerge from the literature.

\textit{i) Single-platform cross-sections dominate.}
Most studies irradiate one board, measure system-level failure rates, and stop there.
Mojica et~al.~\cite{mojica2018radiation}, Wyrwas et~al.~\cite{wyrwas2017proton,wyrwas2019proton}, Santini et~al.~\cite{santini2015reliability}, Roffe et~al.~\cite{roffe2023single}, and Esquer et~al.~\cite{esquer2024single} all report device-level SEE cross-sections, protons, neutrons, or heavy ions, on Raspberry~Pi, Zynq, or Jetson targets running Linux.
These results establish that Linux systems fail under radiation but do not identify where inside the OS failures originate.

\textit{ii) Kernel-level tracing remains rare and incomplete.}
Wilson et al.~\cite{wilson2021neutron} first reported neutron effects on a VexRiscV Linux SoC at device level; their follow-up~\cite{wilson2023post} added post-radiation attribution of failures to hardware subsystems (DDR controller, FPGA interconnect) via kernel-log analysis. Jaksch et al.~\cite{jaksch2022debugging} and Beck et al.~\cite{beck2023evaluation} attributed subsets of observed failures to kernel subsystems on the Raspberry Pi 4B and Zynq UltraScale+ respectively, while Corley et al.~\cite{corley2022accelerated} observed cross-core failure propagation on a Raspberry Pi 3B+. Each study is limited to a single platform and particle type, leaving open whether the observed failure patterns generalize across architectures.

\textit{iii) Cross-architecture comparison lacks OS-level resolution. }
Budroweit et~al.~\cite{budroweit2021risk} surveyed radiation susceptibility across multiple COTS SoCs but without OS-level fault decomposition.
To our knowledge, no prior work combines multi-ISA comparison with kernel-handler fault tracing and propagation-chain reconstruction.

This study addresses that gap with proton irradiation of three platforms spanning two ISAs and two process nodes, a Raspberry Pi Zero 2W (40 nm ARM Cortex-A53), an NXP i.MX 8M Plus (14 nm FinFET ARM Cortex-A53), and an OrangeCrab ECP5 FPGA hosting a VexRiscV soft-core (40 nm RISC-V), combining kernel-subsystem attribution with failure-propagation chain reconstruction.

\begin{table}[!t]
\centering
\caption{Prior Linux-under-radiation studies. $\times$: not addressed; $\checkmark$: addressed; P: partial.}
\label{tab:related_work}
\renewcommand{\arraystretch}{1.2}
\setlength{\tabcolsep}{3.5pt}
\resizebox{\columnwidth}{!}{%
\begin{tabular}{@{}l l p{3.2cm} c c c c@{}}
\toprule
\textbf{Study} & \textbf{Particle} & \textbf{Platform}
  & \rotatebox[origin=c]{90}{\textbf{Cross-arch.}}
  & \rotatebox[origin=c]{90}{\textbf{Kernel trace}}
  & \rotatebox[origin=c]{90}{\textbf{Prop.\ chains}}
  & \rotatebox[origin=c]{90}{\textbf{OS workload}} \\
\midrule
Mojica~\cite{mojica2018radiation}
  & Proton   & RPi Zero, ARM11            & $\times$ & $\times$ & $\times$ & Flight SW \\
Santini~\cite{santini2015reliability}
  & Neutron  & Parallella, Zynq+ARM       & $\times$ & $\times$ & $\times$ & Idle/stress \\
Wyrwas~\cite{wyrwas2017proton}
  & Proton   & RPi 3, Cortex-A53          & $\times$ & $\times$ & $\times$ & Stress \\
Wyrwas~\cite{wyrwas2019proton}
  & Proton   & Zynq-7000, Cortex-A9       & $\times$ & $\times$ & $\times$ & Stress \\
Wilson~\cite{wilson2021neutron}
  & Neutron  & VexRiscV, ECP5 FPGA        & $\times$ & $\times$ & $\times$ & Boot/test \\
Corley~\cite{corley2022accelerated}
  & Neutron  & RPi 3B+, Cortex-A53        & $\times$ & P        & $\times$ & Stress \\
Jaksch~\cite{jaksch2022debugging}
  & Neutron  & RPi 4B, Cortex-A72         & $\times$ & P        & $\times$ & Debug \\
Wilson~\cite{wilson2023post}
  & Neutron  & VexRiscV SoC, ECP5         & $\times$ & $\checkmark$ & $\times$ & Boot/test \\
Roffe~\cite{roffe2023single}
  & Heavy ion & Jetson, ARM+GPU           & $\times$ & $\times$ & $\times$ & Linux+CUDA \\
Beck~\cite{beck2023evaluation}
  & Neutron  & Zynq US+, Cortex-A53       & $\times$ & P        & $\times$ & Stress \\
Esquer~\cite{esquer2024single}
  & Proton, neutron & RPi 4B, Cortex-A72 & $\times$ & $\times$ & $\times$ & Stress \\
Budroweit~\cite{budroweit2021risk}
  & Survey   & Multiple COTS SoCs         & $\checkmark$ & $\times$ & $\times$ & Various \\
\midrule
\textbf{This work}
  & \textbf{Proton}
  & \textbf{RPi Zero 2W, i.MX 8MP,\newline OrangeCrab}
  & $\checkmark$ & $\checkmark$ & $\checkmark$ & \textbf{Stress} \\
\bottomrule
\end{tabular}%
}
\end{table}

\section{Conclusion and Future Work}
\label{sec:conclusions}

Across 133 proton-induced Linux-SEFI events on three COTS platforms, we attributed each failure to its earliest identifiable kernel subsystem and established three principal findings.

First, failure-origin profiles diverge sharply by platform architecture. On both 40 nm platforms, memory-management handlers dominate (39\% on the ARM SoC, 67\% on the RISC-V FPGA), with device drivers contributing an additional 39\% on the ARM platform; the combined memory-facing subsystem share ranges from 67\% to 78\%. On the 14 nm FinFET platform, eMMC-rooted failures filesystem faults (56\%) plus storage-driver errors (34\%) account for 90\% of events, demonstrating that a single peripheral's process node can determine system-level vulnerability regardless of the primary SoC's technology.

Second, the 14 nm FinFET platform exhibits approximately one order of magnitude lower Linux-SEFI cross-sections than both 40 nm devices, although packaging, DRAM-exposure geometry, and irradiation setup differences preclude isolating the FinFET contribution alone.

Third, at 30 MeV the RISC-V soft-core Linux-SEFI cross-section ($7.63 \times 10^{-9}\ \mathrm{cm}^2$) statistically overlaps with the 40 nm ARM result ($6.95 \times 10^{-9}\ \mathrm{cm}^2$) within 95\% confidence intervals, indicating that ISA choice has less influence on system-level vulnerability than peripheral architecture and memory technology at equivalent process nodes.

Failure-propagation analysis reveals that Linux's layered trust model becomes a fault-amplification mechanism under radiation: an eMMC controller SEFI propagated through six kernel abstraction layers into a circular dependency where PID\,1 recovery required the very storage subsystem it was trying to restore. This pattern recurred deterministically three times on DUT\,2, and the firmware-mediated single point of failure identified on the Raspberry Pi platform further confirms that implicit inter-layer trust enables catastrophic single-event cascades. These findings indicate that targeted kernel hardening at critical abstraction boundaries could complement hardware-level redundancy by addressing fault propagation pathways that component-level replication alone does not eliminate.

\section{Future Work}
Our future work will develop a Yocto-based hardened Linux distribution informed by the Linux-SEFI failure data presented in Tables~\ref{tab:sefi_rpi_analysis}, \ref{tab:sefi_imx8mp}, and \ref{tab:sefi_ocv2}, translating observed failure chains into targeted kernel-level mitigations.
We will also co-design the VexRiscv soft-core with selective TMR and ECC at the RTL level, prioritizing the subsystems identified as most SEFI-susceptible.


\newcolumntype{L}{>{\raggedright\arraybackslash}X}

\begin{table*}[!p]
\centering
\caption{Linux-Visible Failure Signatures Recorded During Proton-Irradiation Campaigns on the Raspberry Pi Zero~2W.}
\label{tab:sefi_rpi_analysis}
\footnotesize
\renewcommand{\arraystretch}{1.15}
\setlength{\tabcolsep}{3pt}

\begin{tabularx}{\textwidth}{@{} >{\bfseries}p{1.4cm} L L L @{}}
\toprule
\textbf{Component} & \textbf{Fault Details (Type, Error, \& Subsystems)} & \textbf{Root Cause} & \textbf{Proposed Mitigation} \\
\midrule

\rowcolor{red!25}
\multicolumn{4}{c}{\textbf{Critical-Severity Failures (power-cycle required)}} \\
\addlinespace[1pt]

1.1 \newline Memory (\texttt{mm/}) &
\textbf{Type:} Kernel Oops (Data Abort) --- three instances \newline
\textbf{Error:} \texttt{Unable to handle kernel paging request}; L0/L1 Translation Faults (\texttt{ESR~0x96000004/5}) across VMA teardown, VFS write, and user-return paths. Most severe: recursive fault, oops~[\#2], \texttt{``reboot is needed.''} \newline
\textbf{Subsystems:} \texttt{mm/}, \texttt{fs/}, \texttt{arch/arm64/} &
\textbf{Cause:} Corrupted kernel pointers produced non-canonical or architectural-hole addresses across three subsystems. Patterns are compatible with single-event-upset-class pointer corruption; precise origins not identifiable from logs alone. &
\textbullet~Enable \texttt{CONFIG\_PANIC\_ON\_OOPS=y} with \texttt{kernel.panic=5} to force reboot on first oops. \newline
\textbullet~Validate \texttt{vma->vm\_next} in \texttt{remove\_vma()}: reject bits~[63:48]$\neq$\texttt{0xFFFF}, converting panic into a contained memory leak. \newline
\textbullet~Deploy \texttt{CONFIG\_BCM2835\_WDT=y} (15\,s timeout) to bound recovery time. \\
\midrule

\rowcolor{orange!35}
\multicolumn{4}{c}{\textbf{High-Severity Failures}} \\
\addlinespace[1pt]

2.1 \newline Firmware (\texttt{drivers/\allowbreak firmware/}) &
\textbf{Type:} VideoCore~IV Mailbox Timeout + Persistent Tag Failure \newline
\textbf{Error:} \texttt{Firmware transaction timeout} at \texttt{raspberrypi.c:67}; errno~\texttt{-110}. Separately: \texttt{Request~0x00030046 returned status~0x80000001} (firmware failure flag, 9 consecutive polls). \newline
\textbf{Subsystems:} \texttt{drivers/firmware/\allowbreak raspberrypi.c}, \texttt{raspberrypi\_hwmon}, \texttt{vc4\_hdmi.c} &
\textbf{Cause:} Two unrelated consumers stalled on the shared mailbox path; in a separate boot a specific firmware tag returned persistent error while other tags worked. Precise failure point (hardware, firmware scheduler, or kernel completion path) is not distinguishable from logs alone. &
\textbullet~Disable non-essential VC4 paths: remove \texttt{dtoverlay=vc4-kms-v3d}, set \texttt{hdmi\_ignore\_hotplug=1}. Deploy external I2C thermal sensor. \newline
\textbullet~Patch \texttt{rpi\_firmware\_transaction()} with bounded retry (3 attempts, exponential backoff) and VC4 reset via BCM2710 Power Manager after $N\geq3$ timeouts. \\
\midrule

2.2 \newline Scheduler (\texttt{kernel/\allowbreak sched}) &
\textbf{Type:} Scheduling-While-Atomic BUG + Preempt-Count Underflow WARN \newline
\textbf{Error:} \texttt{BUG: scheduling while atomic: swapper/2/0/\allowbreak 0x080a8000}; \texttt{DEBUG\_LOCKS\_WARN\_ON\allowbreak (val~>~preempt\_count())} at \texttt{preempt\_count\_sub\allowbreak +0x80}. \newline
\textbf{Subsystems:} \texttt{kernel/sched/core.c}, SMP secondary CPU bring-up, MM swap-in &
\textbf{Cause:} CPU~2 idle task reached \texttt{\_\_schedule} with \texttt{preempt\_count}$\neq$0, downstream of an earlier uncaptured fault. Separately, preempt-count underflowed in a well-exercised swap-in path. Control-state corruption is the closest hypothesis; software bug cannot be excluded from single instances. &
\textbullet~Enable \texttt{CONFIG\_DEBUG\_PREEMPT}; configure panic-on-scheduler-BUG for radiation deployments. \newline
\textbullet~Deploy \texttt{bcm2835\_wdt} with short timeout; enable soft-lockup and RCU-stall detectors. \newline
\textbullet~Supervisor: watch dmesg for preempt WARN and \texttt{SIGKILL} offending PID. \\
\midrule

2.3 \newline File System (\texttt{fs/}) &
\textbf{Type:} FAT-fs Dirty Mount (Post-Reset Residue) \newline
\textbf{Error:} \texttt{FAT-fs (sda): Volume was not properly unmounted.} \texttt{dwc\_otg\_hcd\allowbreak ...Unable to get corresponding channel} ($\times$2). \newline
\textbf{Subsystems:} \texttt{fs/fat/inode.c}; \texttt{dwc\_otg} USB HCD; \texttt{sda} &
\textbf{Cause:} Fresh boot observed residue from an unclean prior shutdown. FAT has no journal or metadata checksum, so interrupted updates cannot safely recover. The upstream failure causing the reset cannot be assigned from this evidence alone. &
\textbullet~Replace FAT with ext4 (\texttt{metadata\_csum}, \texttt{journal\_data}) or Btrfs. Enforce \texttt{fsck.mode=force} in kernel command line. \newline
\textbullet~Use read-only rootfs with \texttt{tmpfs} overlays to eliminate transient-to-persistent error propagation. \\
\midrule

\rowcolor{yellow!25}
\multicolumn{4}{c}{\textbf{Medium-to-Low-Severity Failures (Degraded operation)}} \\
\addlinespace[1pt]

3.1 \newline Drivers (\texttt{drivers/\allowbreak gpu/}) &
\textbf{Type:} HDMI Connector Detect Warning \newline
\textbf{Error:} \texttt{WARNING} at \texttt{vc4\_hdmi.c:464}; observed on CPU~0--3 across multiple PIDs and boot sessions; re-fires every $\sim$10.24\,s. \newline
\textbf{Subsystems:} \texttt{vc4\_hdmi.c}, \texttt{drm\_kms\_helper} (\texttt{output\_poll\_execute}) &
\textbf{Cause:} Downstream of the VideoCore mailbox failure in~2.1. Each HDMI check fails when firmware is stuck; deterministic recurrence across boots favours a repeatable post-failure state over per-boot single-event upset. &
\textbullet~Disable DRM output polling: \texttt{echo~0~>~\allowbreak /sys/module/drm\_kms\_helper/\allowbreak parameters/poll}. \newline
\textbullet~Remove \texttt{vc4} module or \texttt{dtoverlay} for headless radiation deployments. \\
\midrule

3.2 \newline Kernel (\texttt{kernel/}) &
\textbf{Type:} Taint Accumulation + Workload Artifact \newline
\textbf{Error:} Taint: \texttt{G~C} $\to$ \texttt{G~W~C} $\to$ \texttt{G~D~C}. OOM by design: \texttt{Killed process~957 (stress-ng) oom\_score\_adj:1000}; kernel resumed normally. \newline
\textbf{Subsystems:} \texttt{kernel/panic.c} (\texttt{add\_taint()}), \texttt{mm/oom\_kill.c} &
\textbf{Cause:} Taint flags recorded worsening failure state: \texttt{W} from warnings~(3.1), \texttt{D} from oopses~(1.1). After \texttt{D}, kernel should be treated as unreliable. OOM is \textit{not} radiation-induced: \texttt{stress-ng} with \texttt{oom\_score\_adj=1000} produces the same kill with the beam off. &
\textbullet~Taint monitor: read \texttt{/proc/sys/kernel/tainted} every second; on bit~9~(\texttt{D}), trigger immediate watchdog reboot. \newline
\textbullet~Per-cgroup \texttt{memory.max}; \texttt{oom\_score\_adj=-1000} on monitor processes. Do \textit{not} set \texttt{panic\_on\_oom}. \\
\midrule

3.3 \newline Memory (DRAM) &
\textbf{Type:} DRAM Single-Bit Upset \newline
\textbf{Error:} \texttt{stress-ng: fail: bit flip at index~4096, bit~7}. Three stressors (gray code, rowhammer, rand-set) independently detected errors at \texttt{0x7facf5a000}, \texttt{0x7fb0112000}. \newline
\textbf{Subsystems:} LPDDR2 (no ECC); \texttt{mm/memory.c} &
\textbf{Cause:} Compare-after-write verifier detected pattern mismatches: silent-data-corruption signature at the application level. Origin not localizable to a specific layer from logs alone. Prior clean run in the same boot weakens the hypothesis of a permanent cell defect. &
\textbullet~Implement kernel-space memory scrubbing (read-verify-rewrite, 10--30\,s interval) with software checksums for single-bit correction. \newline
\textbullet~Enable \texttt{CONFIG\_PAGE\_POISONING=y}; apply software SECDED encoding (12.5\% overhead) for critical data. \\
\bottomrule
\end{tabularx}
\end{table*}

\clearpage
\begin{table*}[!p]
\centering
\caption{Linux SEFI Failure Classification and Resilience Strategies on the NXP i.MX 8M Plus Under Proton Irradiation.}
\label{tab:sefi_imx8mp}
\footnotesize
\renewcommand{\arraystretch}{1.15}
\setlength{\tabcolsep}{3pt}

\begin{tabularx}{\textwidth}{@{} >{\bfseries}p{1.25cm}
>{\raggedright\arraybackslash}X
>{\raggedright\arraybackslash}X
>{\raggedright\arraybackslash}X @{}}
\toprule
\textbf{Component} & \textbf{Fault Details: Type, Error, \& Subsystems} & \textbf{Root Cause and Cascading Impact} & \textbf{Proposed Mitigation} \\
\midrule

\rowcolor{red!25}
\multicolumn{4}{c}{\textbf{Critical Severity Failures (Power Cycle Required)}} \\
\addlinespace[1pt]

1.1 \newline Drivers (\texttt{drivers/\allowbreak mmc/}) &
\textbf{Type:} eMMC or SDHCI Controller SEFI \newline
\textbf{Error:} \texttt{mmc2: Card stuck being busy! \_\_mmc\_poll\_for\_busy}; \texttt{cache flush error \char45{}110}; \texttt{HW reset card, got error \char45{}110}; \texttt{mmcblk2: recovery failed!} \newline
\textbf{Code:} \texttt{WARNING} at \texttt{sdhci\_send\_command\allowbreak +0x4fc/0xeac}, \texttt{lr: sdhci\_send\_command\allowbreak \_retry+0x40/0x130} \newline
\textbf{Subsystems:} \texttt{sdhci\_esdhc\_imx.c}, USDHC \texttt{0x30b60000}; \texttt{mmc/core/}, including \texttt{core.c}, \texttt{block.c}, and \texttt{mmc.c} &
\textbf{Cause:} SEFI in the USDHC controller or eMMC internal controller, likely 28 to 40~nm, locked the eMMC in a permanent busy state during CMD6, \texttt{\_\_mmc\_switch}. Cache flush and \texttt{RST\_n} both returned \texttt{\char45{}ETIMEDOUT}, indicating a deep SEFI requiring a full VCC power cycle. Occurred 3 times. \newline
\textbf{Impact:} All block I/O returns \texttt{\char45{}EIO}; all downstream failures, 1.2, 2.1, 2.2, 3.1, and 3.2, cascade from this single SEFI. &
\textbullet~Implement eMMC VCC power cycling via PMIC, PCA9450, GPIO, bypassing insufficient \texttt{RST\_n}. Register a custom \texttt{mmc\_hw\_reset} that cuts VMMC for 10~ms per JEDEC \texttt{t\_VCCQ\_stable}. \newline
\textbullet~Move the root filesystem to QSPI NOR or NFS root. Confine eMMC to noncritical storage with checksummed writes. \\
\midrule

1.2 \newline Kernel (\texttt{kernel/}) \newline and Init &
\textbf{Type:} Multi Process SIGILL or SIGSEGV plus PID~1 Freeze \newline
\textbf{Error:} \texttt{systemd[1]: Caught <SEGV>, core dump failed}; \texttt{Freezing execution.} Six processes crashed: 3$\times$SIGILL, \texttt{sh}, \texttt{dbus\_daemon}, and \texttt{avahi\_daemon}; and 3$\times$SIGSEGV, \texttt{ModemManager}, \texttt{systemd}, and \texttt{systemd\_journald}. \newline
\textbf{Subsystems:} \texttt{arch/arm64/kernel/traps.c}, \texttt{kernel/signal.c}, \texttt{mm/filemap.c}, systemd crash handler &
\textbf{Cause:} eMMC I/O errors from Failure~1.1 caused the page cache to serve corrupted instruction pages. Processes sharing text mappings, likely \texttt{libc.so.6}, executed invalid opcodes and raised SIGILL. systemd caught SIGSEGV, failed core dump because the filesystem was read only, and entered infinite \texttt{pause()}. PID~1 frozen means system dead. \newline
\textbf{Impact:} D Bus IPC dead; journal corrupted; system unrecoverable without hardware reset. &
\textbullet~Deploy dm verity, \texttt{CONFIG\_DM\_VERITY=y}, to verify block integrity via Merkle tree, converting silent corruption into clean \texttt{\char45{}EIO}. \newline
\textbullet~Patch the systemd crash handler to call \texttt{reboot(RB\_AUTOBOOT)} via PSCI \texttt{SYSTEM\_RESET} instead of \texttt{freeze()}, bypassing filesystem dependent reboot. \\
\midrule

\rowcolor{orange!35}
\multicolumn{4}{c}{\textbf{High Severity Failures (Graceful Linux Reboot Required)}} \\
\addlinespace[1pt]

2.1 \newline File System (\texttt{fs/}) &
\textbf{Type:} EXT4 or JBD2 Journal Abort plus Read Only Remount \newline
\textbf{Error:} \texttt{ext4\_end\_bio:344: I/O error~10}; \texttt{Aborting journal on mmcblk2p1\char45{}8}; \texttt{JBD2: Error \char45{}5 updating journal superblock}; \texttt{Remounting filesystem read only}. Occurs 3 times. \newline
\textbf{Subsystems:} \texttt{fs/ext4/}, including \texttt{page\_io.c} and \texttt{super.c}; \texttt{fs/jbd2/journal.c}; \texttt{block/blk\_core.c} &
\textbf{Cause:} Cascade from Failure~1.1. eMMC \texttt{\char45{}EIO} propagates through the block layer to EXT4 BIO completion. JBD2 cannot write its superblock, \texttt{\char45{}EIO}, and aborts the journal. EXT4 remounts read only. \newline
\textbf{Impact:} \texttt{chmod /var/lib/docker: read only file system}; \texttt{overlayfs: upper fs is r/o}, causing container storage failure; superblock write failure at sector~8192. &
\textbullet~Mount root with \texttt{errors=panic} plus \texttt{kernel.panic=1} to force immediate reboot on first I/O error. \newline
\textbullet~Use RAM backed overlayfs root, with eMMC as read only lower layer and \texttt{tmpfs} as upper layer. Cached pages survive eMMC failure. \\
\midrule

2.2 \newline Drivers and Runtime &
\textbf{Type:} Docker SIGBUS Double Panic \newline
\textbf{Error:} \texttt{dockerd[783]: [signal SIGBUS: bus error code=0x2 addr=0xaaaaaadc1d3c0 pc=0xaaaaaabb25668]}; \texttt{panic during panic}; second SIGBUS at same pc. Docker restarts 4 times, then \texttt{service start limit hit}. \newline
\textbf{Subsystems:} \texttt{mm/memory.c}, \texttt{mm/filemap.c}, \texttt{arch/arm64/mm/fault.c}, Go \texttt{runtime} &
\textbf{Cause:} Cascade from Failure~1.1. \texttt{dockerd} demand pages text from eMMC; page fault returns \texttt{\char45{}EIO} via \texttt{ext4\_readpage()}, delivered as SIGBUS, \texttt{BUS\_ADRERR}, code=0x2. The Go runtime panic handler accesses more unmapped pages, triggering a second SIGBUS. \newline
\textbf{Impact:} Docker daemon dead; all containers lose management; subsequent restarts fail because the filesystem is read only. &
\textbullet~Pin \texttt{dockerd} in memory via \texttt{mlockall(MCL\_CURRENT|MCL\_FUTURE)} at startup to prevent demand paging from eMMC. \newline
\textbullet~Use a lightweight static supervisor, locked in RAM and independent of Docker, for critical workloads. \\
\midrule

\rowcolor{yellow!25}
\multicolumn{4}{c}{\textbf{Medium to Low Severity Failures (Degraded Operation)}} \\
\addlinespace[1pt]

3.1 \newline Init (\texttt{systemd}) &
\textbf{Type:} Reboot Target Load Failure \newline
\textbf{Error:} \texttt{Failed to enqueue replace irreversibly job for reboot.target: Input/output error}. Occurs 4 times. \newline
\textbf{Subsystems:} systemd unit loader, filesystem read path &
\textbf{Cause:} Cascade from Failures~1.1 and 2.1. systemd cannot load \texttt{reboot.target} from read only or \texttt{\char45{}EIO} eMMC. The software reboot path is broken. Combined with PID~1 freeze from Failure~1.2, only hardware watchdog recovery remains. \newline
\textbf{Impact:} System cannot self recover; indefinite hang until watchdog or external power cycle. &
\textbullet~Use a kernel module that triggers \texttt{emergency\_restart()} via PSCI on sustained I/O errors, bypassing systemd. \newline
\textbullet~Cache critical unit files in \texttt{tmpfs} at early boot via \texttt{systemd.unit\_path=/run/systemd/units}. \\
\midrule

3.2 \newline File System (\texttt{fs/}) &
\textbf{Type:} FAT fs Dirty Mount, USB Storage \newline
\textbf{Error:} \texttt{FAT fs (sda1): Volume was not properly unmounted. Some data may be corrupt.} Occurs 3 times. \newline
\textbf{Subsystems:} \texttt{fs/fat/inode.c}, USB storage \texttt{sda1} &
\textbf{Cause:} Consequence of radiation induced crashes, Failures~1.1 and 1.2. Each crash bypassed shutdown, leaving FAT dirty flags set. FAT lacks journaling, checksums, and write ordering. \newline
\textbf{Impact:} No automatic recovery without \texttt{fsck}. &
\textbullet~Replace FAT with ext4 using \texttt{metadata\_csum}, or use F2FS. Enforce \texttt{fsck.mode=force} at boot. \newline
\textbullet~Mount USB with the \texttt{sync} option to minimize the data at risk window. \\

\bottomrule
\end{tabularx}
\end{table*}

\clearpage
\begin{table*}[!p]
\centering
\caption{Linux-SEFI Failure Classification and Resilience Strategies on the OrangeCrab FPGA Under Proton Irradiation.}
\label{tab:sefi_ocv2}
\footnotesize
\renewcommand{\arraystretch}{1.15}
\setlength{\tabcolsep}{3pt}

\begin{tabularx}{\textwidth}{@{} >{\bfseries}p{1.4cm} L L L @{}}
\toprule
\textbf{Component} & \textbf{Fault Details (Type, Error, \& Subsystems)} & \textbf{Root Cause \& Cascading Impact} & \textbf{Proposed Mitigation} \\
\midrule

\rowcolor{red!25}
\multicolumn{4}{c}{\textbf{Critical-Severity (power-cycle required)}} \\
\addlinespace[1pt]

1.1 \newline Memory (\texttt{mm/}) &
\textbf{Type:} NULL Ptr Deref / Kernel Paging Failure / SIGSEGV \newline
\textbf{Error:} \texttt{Unable to handle kernel NULL pointer dereference at virtual address 00000000}; \texttt{kernel paging request at virtual address c00240c0}; Segmentation Fault. \newline
\textbf{Registers:} \texttt{sepc:355b4e14}, \texttt{scause:0000000d}, \texttt{sbadaddr:00000000} \newline
\textbf{Subsystems:} \texttt{riscv/mm/fault.c}, \texttt{litex/.../vexriscv/core.py}, \texttt{litex/.../ecc.py} &
\textbf{Cause:} SEU in DDR3L or VexRiscV registers corrupted a PTE, resolving to NULL (\texttt{0x00000000}) or invalid address (\texttt{0xc00240c0}), triggering store/AMO page fault (\texttt{scause=0xd}). No ECC on DDR3L or FPGA block RAM leaves all kernel data structures exposed. \newline
\textbf{Impact:} \texttt{CPU:0 PID:843 Comm:sh Not tainted 5.0.13 \#1}; kernel crash or \texttt{SIGSEGV}; cascading page table corruption. &
\textbullet~Enable SECDED ECC in LiteDRAM~\cite{litedram_ecc_2024}; add ECC for critical VexRiscV registers via RTL modifications. \newline
\textbullet~Apply DMR/TMR across LiteX SoC (CPUs, memory controllers, buses). See Tables~\ref{tab:sefi_rpi_analysis},~\ref{tab:sefi_imx8mp}. \\
\midrule

1.2 \newline Kernel &
\textbf{Type:} Interrupt Handler Fatal Exception \newline
\textbf{Error:} \texttt{Kernel panic -- not syncing: Fatal exception in interrupt} \newline
\textbf{Subsystems:} \texttt{kernel/traps.c}, \texttt{litex/.../vexriscv/core.py}, \texttt{litex/.../watchdog.py} &
\textbf{Cause:} SEFI corrupted a VexRiscV interrupt controller or PLIC register, causing invalid ISR address during dispatch. ISRs execute non-preemptibly; any exception escalates to non-recoverable panic. PLIC registers in SRAM-based FPGA fabric lack upset protection. \newline
\textbf{Impact:} Immediate system halt; no graceful shutdown or logging; requires full power cycle. &
\textbullet~Add parity/TMR on PLIC interrupt vector table and pending-interrupt bitmap via LiteX RTL. Deploy LiteX watchdog (\texttt{watchdog.py}) for automated reset on interrupt-context hangs. \newline
\textbullet~Enable FPGA configuration scrubbing to correct SRAM bitstream upsets corrupting synthesized PLIC logic. \\
\midrule

\rowcolor{orange!35}
\multicolumn{4}{c}{\textbf{High-Severity (reboot required)}} \\
\addlinespace[1pt]

2.1 \newline Memory (\texttt{mm/}) &
\textbf{Type:} Corrupted Page Map / Bad Page State \newline
\textbf{Error:} \texttt{BUG: Bad page map in process md5sum pte:10444cd7 pmd:11bb7001}; \texttt{bad pte / nonzero mapcount}; \texttt{Bad page state in process kthread pfn:46f5a}. \newline
\textbf{Call Trace:} \texttt{print\_bad\_pte} $\rightarrow$ \texttt{unmap\_page\_range} $\rightarrow$ \texttt{exit\_mmap} $\rightarrow$ \texttt{do\_exit} $\rightarrow$ \texttt{ret\_from\_syscall}. \newline
\textbf{Subsystems:} \texttt{mm/(fault.c, memory.c, pgtable.c, page\_alloc.c)}, \texttt{litex/.../vexriscv/core.py}, \texttt{liteDRAM/.../controller.py} &
\textbf{Cause:} SEU corrupted PTE (\texttt{0x10444cd7}) or PMD (\texttt{0x11bb7001}) in unprotected DDR3L, detected during \texttt{unmap\_page\_range()} at process exit. Nonzero \texttt{mapcount} on freed page indicates radiation disrupted reference counting. May not manifest until process exit, allowing silent propagation. \newline
\textbf{Impact:} Corrupted page tables for \texttt{md5sum}/\texttt{kthread}; use-after-free risk; cascading panic if corrupted pages reallocated to kernel structures. &
\textbullet~Redundant PTE/PMD copies in LiteX SoC; parity checks on page table entries with automatic TLB invalidation upon repair. \newline
\textbullet~Harden LiteDRAM controller with ECC to correct single-bit upsets in page table regions before kernel propagation. \\
\midrule

\rowcolor{yellow!25}
\multicolumn{4}{c}{\textbf{Medium/Low-Severity (degraded operation or warning only)}} \\
\addlinespace[1pt]

3.1 \newline Kernel &
\textbf{Type:} Kernel Oops \newline
\textbf{Error:} \texttt{Oops [\#00001]}; subsequently: \texttt{Unable to handle kernel NULL pointer dereference at virtual address 0000}. \newline
\textbf{Subsystems:} \texttt{kernel/traps.c}, \texttt{litex/.../vexriscv/core.py} &
\textbf{Cause:} Non-fatal trap in VexRiscV triggered an Oops, killing only the offending process but leaving the kernel in inconsistent state. On single-core VexRiscV, no redundant core maintains services during recovery. \newline
\textbf{Impact:} Degraded operation; subsequent faults more likely; diagnostic reliability compromised. &
\textbullet~Node-level TMR: three VexRiscV instances on 85K-LUT ECP5 (or three OrangeCrab boards) with majority voting, analogous to SpaceX Falcon~9~\cite{Rajadurai2020}. \newline
\textbullet~Process-level DMR/TMR with lockstep for critical processes. See Table~\ref{tab:sefi_rpi_analysis}, Section~3.1. \\
\midrule

3.2 \newline Memory (\texttt{mm/}) &
\textbf{Type:} User-Space SIGSEGV / SysFS Corruption \newline
\textbf{Error:} \texttt{md5sum[1140]: unhandled signal 11 at 0x3ffcfb60 in ld-2.29.so}; \texttt{find[123]: signal 11 at 0x9fe0bc0f in busybox}; SysFS read failure on \texttt{/sys/kernel/slab/kmalloc-8/}. \newline
\textbf{Registers:} \texttt{scause:0000000d}, \texttt{sbadaddr:9feb0bcf} \newline
\textbf{Subsystems:} \texttt{kernel/signal.c}, \texttt{mm/slub.c}, \texttt{litex/.../csr\_bus.py} &
\textbf{Cause:} \texttt{SEGV\_MAPERR}, processes accessed unmapped addresses. Faults in shared library region suggest SEU corruption of \texttt{ld-2.29.so} GOT/relocation tables. SysFS failure points to corrupted SLUB metadata. CSR bus upsets likely delivered corrupted data to kernel. \newline
\textbf{Impact:} Multiple process terminations; reduced observability via corrupt SysFS; shared library corruption impacts all linked processes. &
\textbullet~Periodic checksums of critical \texttt{.so} files; recover via \texttt{dlclose()}/\texttt{dlopen()} or \texttt{LD\_PRELOAD} fallback. \newline
\textbullet~Harden LiteX CSR bus with parity/ECC on data transactions to block corrupt reads before kernel propagation. \\
\bottomrule
\end{tabularx}
\end{table*}

\clearpage
\begin{IEEEbiography}[{\includegraphics[width=1in,height=1.25in,clip,keepaspectratio]{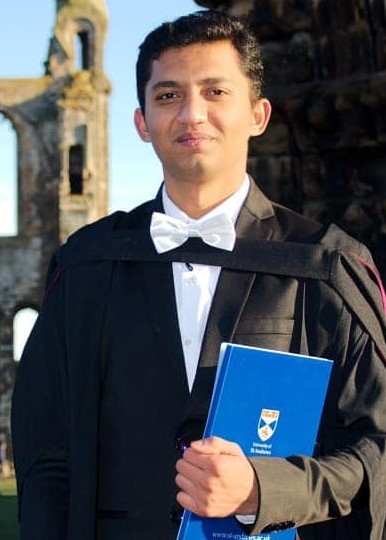}}]{Saad Memon} received the Erasmus Mundus European Double M.Sc. degree in Dependable Software Systems from the University of St. Andrews, UK, in 2019, with a specialization in Artificial Intelligence. He is currently pursuing a Ph.D. as a Doctoral Researcher with the SERVAL group at the Interdisciplinary Centre for Security, Reliability and Trust (SnT), University of Luxembourg. His research interests include hardware-software co-design for radiation-resilient space systems, focusing on leveraging COTS SoCs to develop robust computer architectures and operating systems for AI/ML applications.
\end{IEEEbiography}

\begin{IEEEbiography}[{\includegraphics[width=1in,height=1.25in,clip,keepaspectratio]{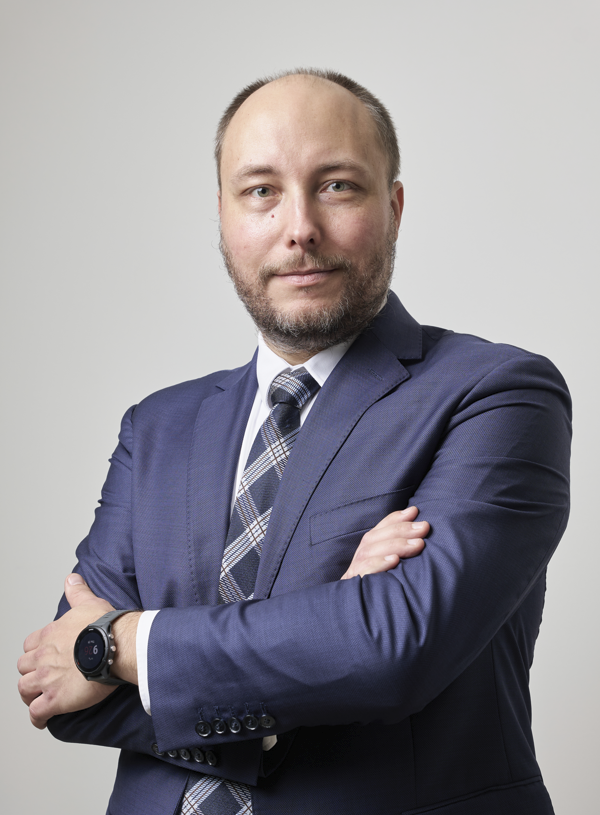}}]{Rafal Graczyk} received the B.Sc. and M.Sc. degrees in electronics and computer engineering and the Ph.D. degree in electronics from Warsaw University of Technology, in 2005, 2009, and 2017, respectively. He has more than 15 years of experience in dependable systems research and development, with a focus on resilience and radiation effects in computer systems. He has been active in wide range of projects, conducted with the Space Research Center of the Polish Academy of Sciences, Interdisciplinary Centre for Security, Reliability and Trust at University of Luxembourg, and, as of recently, companies within European space sector, where he assumes technical leadership roles.
\end{IEEEbiography}

\begin{IEEEbiography}[{\includegraphics[width=1in,height=1.25in,clip,keepaspectratio]{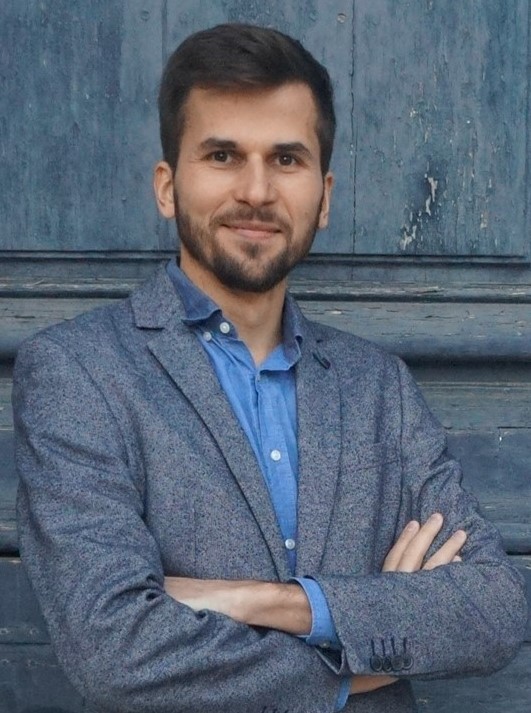}}]{Tomasz Rajkowski} graduated from the Electronics Department at Warsaw University of Technology. From 2011 to 2017, he worked at the Space Research Center of the Polish Academy of Sciences as an electronics test engineer (BRITE-PL project) and later as a hardware designer and project manager (ESA OPS-SAT project). Between 2018 and 2020, he completed his PhD at the University of Montpellier, conducting research at 3D-Plus on the radiation qualification of space systems. Currently, he is a postdoctoral researcher at the National Centre for Nuclear Research, specializing in radiation hardness assurance and serving as an engineer and consultant for several national and European space projects.
\end{IEEEbiography}

\begin{IEEEbiography}[{\includegraphics[width=1in,height=1.25in,clip,keepaspectratio]{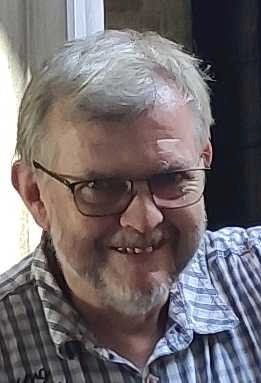}}]{Jan Swakoń} Ph.D., D.Sc., professor of IFJ PAN, is Head of the Department of Radiation Research and Proton Radiotherapy at The H. Institute of Nuclear Physics, Polish Academy of Sciences in Krakow, Poland. He was responsible for building the proton eye radiotherapy facilities at the IFJ PAN and preparing the first proton ocular radiotherapy in Poland. His main scientific activity focuses on new methods of proton radiotherapy, dosimetry, medical and radiation physics including proton and gamma irradiations for space and high-energy physics applications.
\end{IEEEbiography}

\begin{IEEEbiography}[{\includegraphics[width=1in,height=1.25in,clip,keepaspectratio]{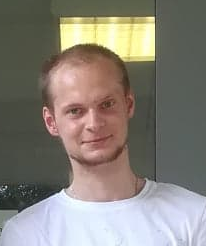}}]{Damian Wrobel} graduated in technical physics from AGH University of Science and Technology. Currently, he is a PhD student at the Institute of Nuclear Physics, Polish Academy of Sciences. His doctoral research is focused on dosimetry for proton FLASH radiotherapy. In particular, he is developing a proton beam forming system and applying radioluminescent crystals for dose determination in ultra-high dose rate proton irradiations.
\end{IEEEbiography}

\begin{IEEEbiography}[{\includegraphics[width=1in,height=1.25in,clip,keepaspectratio]{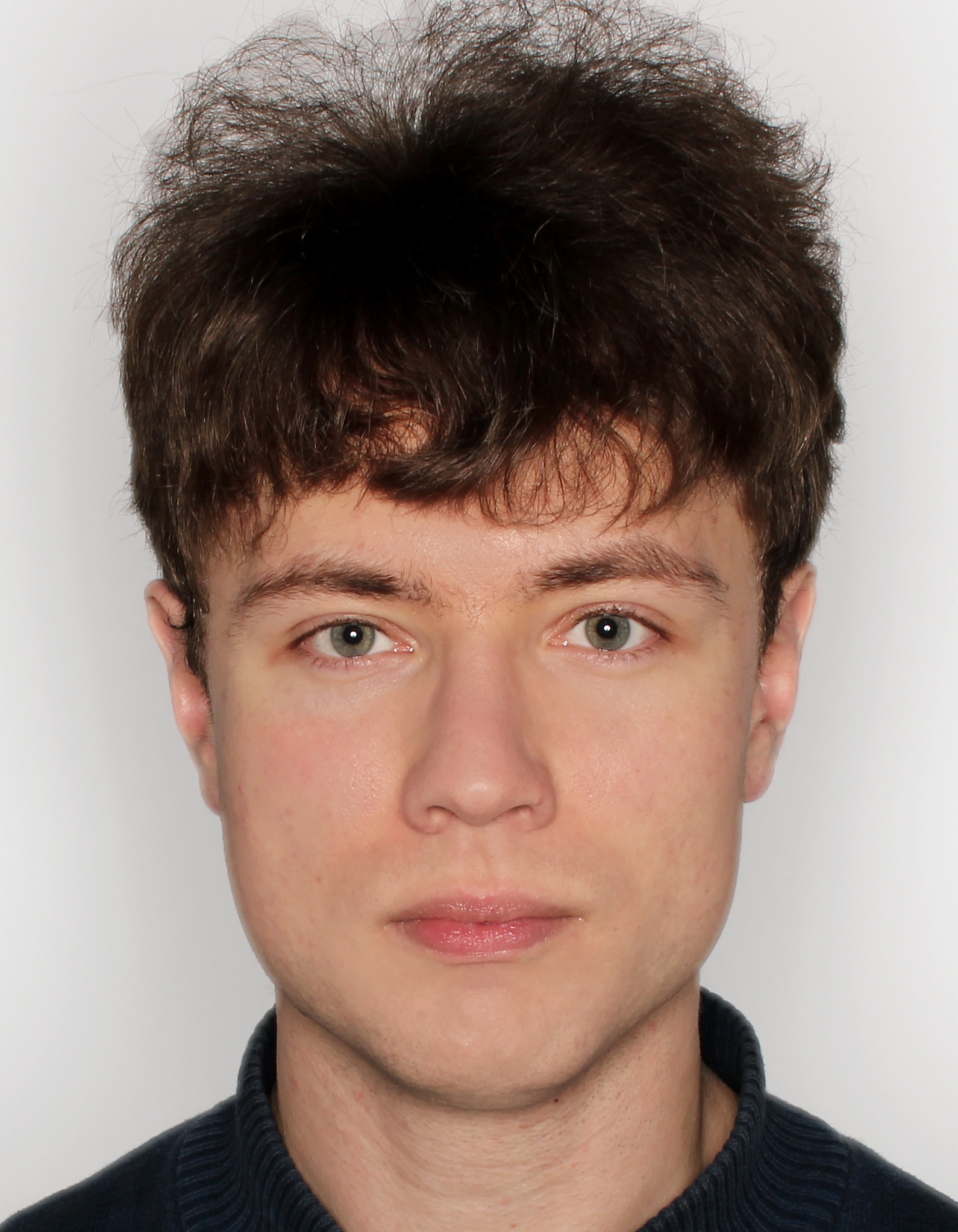}}]{Sebastian Kusyk} is a PhD student at the Institute of Nuclear Physics, Polish Academy of Sciences, since 2022. Subject of doctoral thesis: Proton beam dosimetry for electronics radiation hardness tests (development of proton irradiation station at AIC-144 cyclotron, beam forming, and dosimetry). Proton and gamma (Co-60) irradiations of electronics, new passive dosemeters (i.e., TLD, OSL), biological and other samples. Graduate of Advanced Materials and Nanotechnology at Jagiellonian University.
\end{IEEEbiography}

\begin{IEEEbiography}[{\includegraphics[width=1in,height=1.25in,clip,keepaspectratio]{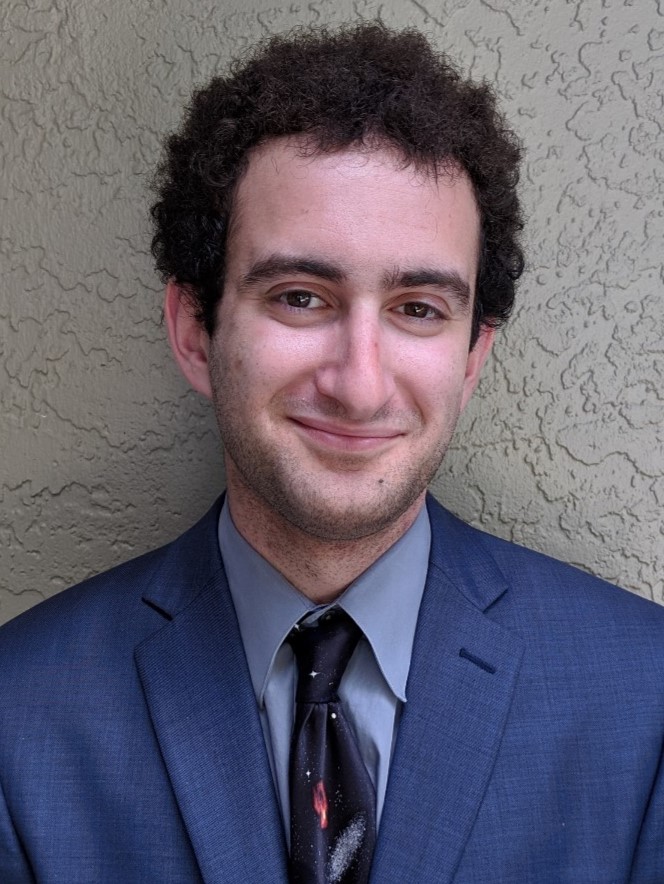}}]{Seth Roffe} received a B.S. degree in physics, astronomy, and mathematics, and M.S. and Ph.D. degrees in electrical and computer engineering from the University of Pittsburgh, Pittsburgh, PA, USA. He is currently a Civil Servant with NASA Goddard Space Flight Center and a part of the Radiation Effects and Analysis Group, primarily performing research on the radiation effects of complex devices and machine-learning applications. His research interests involve resilience in sensor processing, including data reliability and error classification in novel sensors and processors or accelerators.
\end{IEEEbiography}

\begin{IEEEbiography}[{\includegraphics[width=1in,height=1.25in,clip,keepaspectratio]{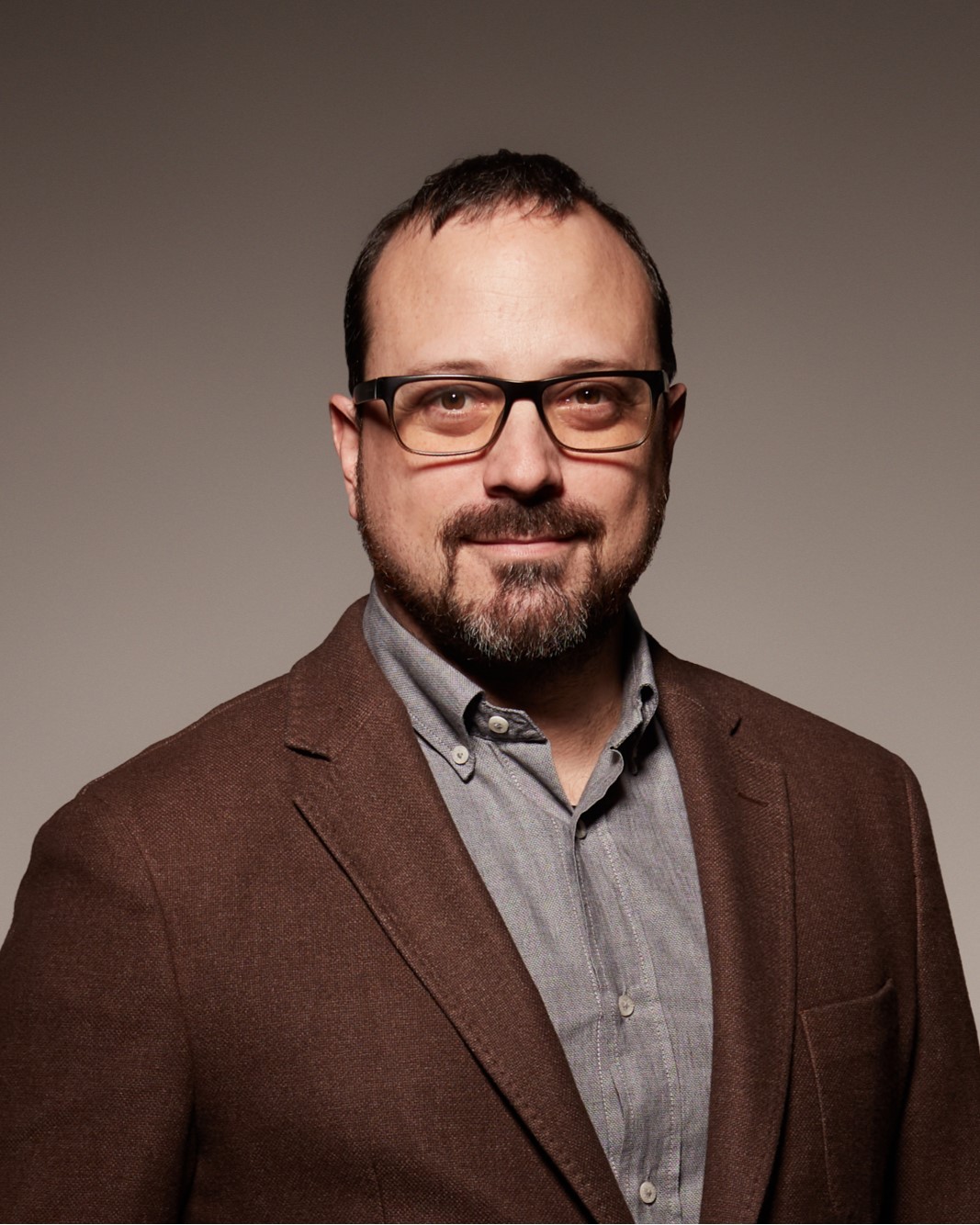}}]{Mike Papadakis} is an associate professor at the Interdisciplinary Center for Security, Reliability and Trust (SnT) of the University of Luxembourg. He received a Ph.D. degree in Computer Science from the Athens University of Economics and Business in 2011. His research interests include software testing, static analysis, prediction modeling, and search-based software engineering. He is best known for his work on Mutation Testing, for which he has been awarded the IEEE TCSE Rising Star Award 2020. He has been awarded several ACM SIGSOFT Distinguished Paper and Artifact Awards and a Facebook Research Award (2019). He is the general (co-)chair of the 37th IEEE International Conference on Software Maintenance and Evolution (ICSME 2021); he has been elected to the steering committees of the IEEE International Conference on Software Testing (ICST), the Symposium on Search-Based Software Engineering (SSBSE) and serves on the editorial and review boards of international Software Engineering journals (STVR, Empirical Software Engineering, ACM Transactions on Software Engineering and Methodology). He has (co-)authored more than 80 publications in international peer-reviewed conferences and journals. His work has been supported by Facebook, FNR, CETREL (SIX group company), BGL (BNP Paribas), Microsoft, and PayPal.
\end{IEEEbiography}

\bibliographystyle{IEEEtran}
\bibliography{references}
\end{document}